\documentclass[aps,prb,superscriptaddress,tightenlines,twocolumn,floatfix,notitlepage]{revtex4-1}
\usepackage{graphicx}
\usepackage[english]{babel}
\usepackage{amsmath,amsthm,mathrsfs,amsfonts,amssymb}   
\usepackage{tensor}
\usepackage{graphicx}  
\usepackage{subfigure}
\usepackage{float}
\usepackage[colorlinks,linkcolor=blue,citecolor=blue,urlcolor=blue]{hyperref}
\usepackage{cleveref}
\crefname{equation}{Eq.}{Eq.}
\crefname{figure}{Fig.}{Fig.}
\crefname{table}{Tab.}{Tab.}
\crefname{section}{Sec.}{Sec.}

\usepackage[T1]{fontenc}
\usepackage{mathptmx}
\DeclareMathAlphabet{\mathcal}{OMS}{cmsy}{m}{n}
\DeclareSymbolFont{Letters}{OML}{cmm}{m}{it}
\DeclareMathSymbol{\psi}{\mathalpha}{Letters}{32}
\DeclareMathSymbol{\Psi}{\mathalpha}{Letters}{9}

\newcommand{\vk}{\mathbf{k}}

\newcommand{\be}{\begin{eqnarray}}
\newcommand{\ee}{\end{eqnarray}}

\newcommand{\beq}{\begin{equation}}
\newcommand{\eeq}{\end{equation}}
\newcommand{\ben}{\begin{align}}
\newcommand{\een}{\end{align}}
\newcommand{\bea}{\begin{aligned}}
\newcommand{\eea}{\end{aligned}}
\newcommand{\bes}{\begin{subequations}}
\newcommand{\ees}{\end{subequations}}
\newcommand{\bew}{\begin{widetext}}
\newcommand{\eew}{\end{widetext}}

\begin{document}

\title{Exact Solution to Sodium-Iridate-BCS-Hubbard Model along the Symmetric Line: Non-trivial topology in the ferromagnetic order}
\author{Shihao Bi}
\email{bishihao@stu.scu.edu.cn}
\affiliation{College of Physics, Sichuan University, 610064, Chengdu, People's Republic of China\\
and Key Laboratory of High Energy Density Physics and Technology of Ministry of Education, Sichuan University, 610064,
Chengdu, People's Republic of China}

\author{Yan He}
\email{heyan_ctp@scu.edu.cn}
\affiliation{College of Physics, Sichuan University, 610064, Chengdu, People's Republic of China\\
and Key Laboratory of High Energy Density Physics and Technology of Ministry of Education, Sichuan University, 610064,
Chengdu, People's Republic of China}

\author{Peng Li}
\email{lipeng@scu.edu.cn}
\affiliation{College of Physics, Sichuan University, 610064, Chengdu, People's Republic of China\\
and Key Laboratory of High Energy Density Physics and Technology of Ministry of Education, Sichuan University, 610064,
Chengdu, People's Republic of China}

\begin{abstract}
We study the sodium-iridates model on the honeycomb lattice with both BCS pairing potential and Hubbard interaction term.
It is shown that this model can be exactly solved with appropriate choices of amplitude of pairing gaps, where the interacting terms are transformed to external field terms. The band structures of these exact solutions on both torus and cylinder geometry are discussed in great details. It is found that the ground state assumes an anti-ferromagnetic configuration, which breaks the time reversal symmetry spontaneously and renders the superconductor topologically trivial.
On the other hand, the nontrivial topology is preserved with ferromagnetic configuration and can be characterized by the isospin Chern number.
\end{abstract}

\maketitle

\section{Introduction}
\label{sec:intro}

Topological quantum matters have been a central topic in the area of condensed matter physics for the last decade \cite{Kane-review,Zhang-review}. The topology of gapped non-interacting fermionic systems has been classified according to three types of discrete symmetries, which leads to the famous ten-fold way classification scheme \cite{Ryu2008,Kitaev-AIP}. The scheme has also been generalized and applied to systems with gapless dispersion and spatial symmetries \cite{Chiu2016:rmp}. On the other hand, the topological properties of interacting fermionic systems have always been intriguing topics, especially in the strongly correlated limit. For example, the pioneer works of Kitaev and coworkers showed that the Hubbard interaction can reduce the $\mathbb{Z}$ classification of topological superconductors to a finite group such as $\mathbb{Z}_8$ \cite{Kitaev-1,Kitaev-2,Yao}. More general considerations of interacting fermionic systems lead to the concept of symmetry protected topological state \cite{Wen-SPT}, which is still under active study.

In some recent works \cite{TKNg2018:prl, TKNg2019:prb, Miao2017:prb, Miao2017:prl, Miao2019:prb, Ezawa2017:prb, Ezawa2018:prb}, it is proposed that certain BCS superconductor with Hubbard interactions can be exactly solved when the amplitude of pairing potentials are tuned to be equal to the hopping constants, i.e. along a symmetric line in the parameter space. This type of exact solution opens up a new approach to study the topological superconductors with arbitrary interaction strength. The mechanism behind these exact solutions strongly resembles the exact solution of Kitaev spin liquid model on the Honeycomb lattice \cite{Kitaev-model}. When the BCS model is expressed in terms of Majorana fermions, half of them has zero kinetic term along the symmetric line. This gives rise to infinitely many conserved quantities and also transforms the Hubbard interacting terms into simple quadratic terms of fermions. In Ref. \cite{Ezawa2018:prb}, Ezawa carried out a detailed study of the BCS superconductor based on the Kane-Mele model with Hubbard interaction. It is known that Kane-Mele model is proposed for graphene, which does not has strong enough spin orbital coupling to become topologically nontrivial. In this paper, we propose the topological superconductor based on the sodium-iridate model \cite{Shitade2009:prl,Ruegg2012:prl}, which possesses a stronger spin orbital coupling. This model can also be exactly solved along the symmetric line when Hubbard interacting terms are considered. It will enrich the family of such exactly solvable models and provide a valuable alternative for exploring the nontrivial topology in these interacting systems.

This paper is organized as follows. In \cref{sec:mod}, we introduce the model Hamiltonian and discuss its symmetry and topological classification. We also propose a possible experimental construction of the system. Then in \cref{sec:SIBCS}, the non-interacting limit is examined in detail. The band structures are displayed in both torus and cylinder geometry. The isospin Chern number is computed to confirm the bulk-edge correspondence in this model. With the help of isospin Chern number, we also obtain the phase diagram. Next we introduce the on-site Hubbard interactions and reveal the exact solvability in \cref{sec:ESU}. By expressing the fermions in terms of Majorana fermions under the perfect flat band condition, we demonstrate that one of the species of the Majorana fermions is decoupled, which makes the interacting terms quadratic. We analyze the band structure and topological properties of the exact solution in two special configurations. Finally we make a conclusion in \cref{sec:con}. We will use the convention of $\hbar=1$ in the rest of this paper.

\section{Model Hamiltonian}
\label{sec:mod}

\begin{figure}
\centering
\includegraphics[width=0.6\columnwidth]{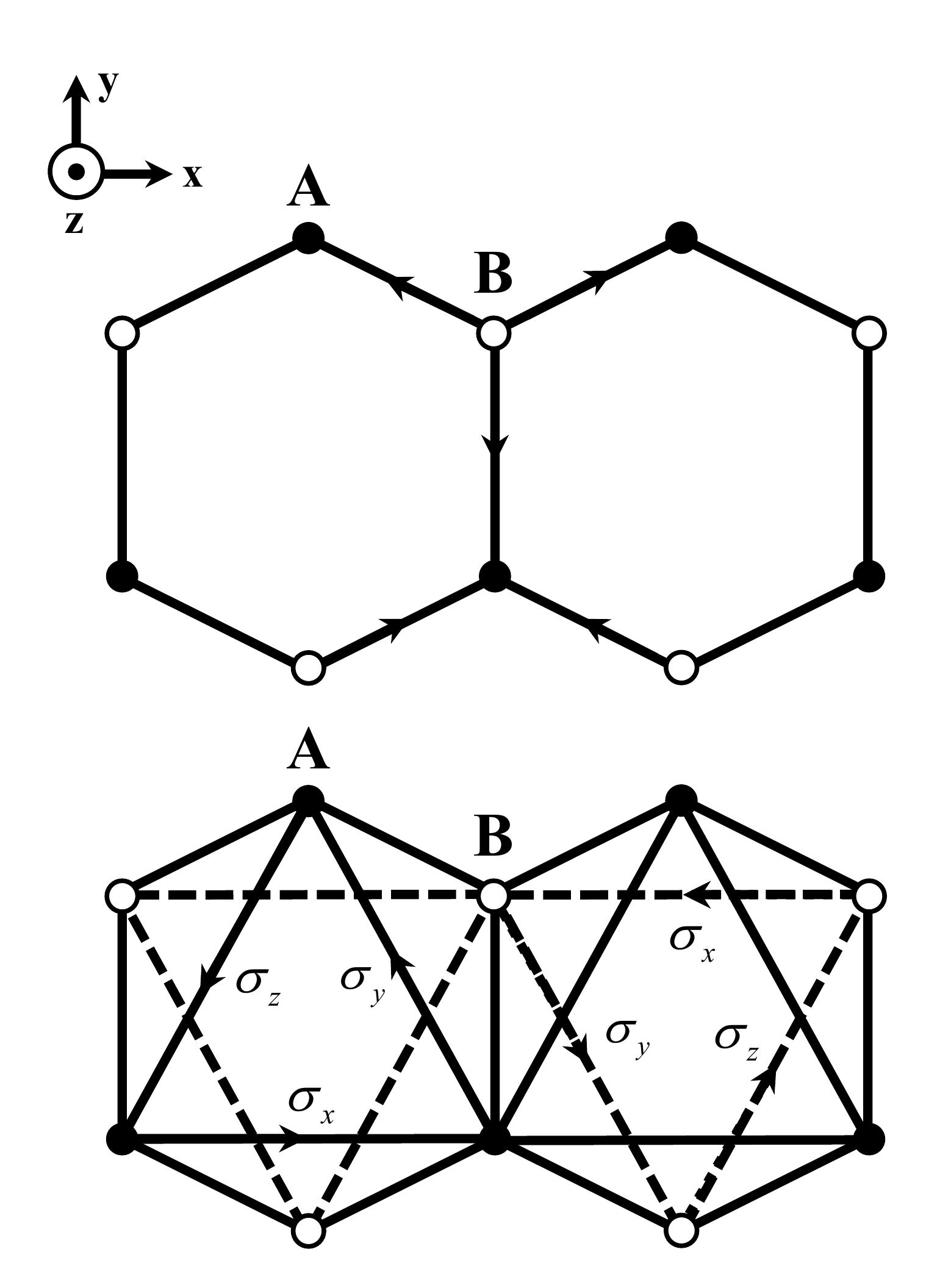}
\caption{Upper panel: The plus (minus) signs of superconducting pairing between the nearest neighbor sites are shown by the forward (backward) direction of the arrows. 
Lower panels: The anisotropic pairing term between the next nearest neighbor sites for sublattices are shown by solid and dashed lines with labels.}
\label{fig:sketch}
\end{figure}

The model Hamiltonian focused in this work consists of three terms,
\begin{equation}
H = H_{\text{SI}} + H_{\text{pair}} + H_{\text{int}},
\label{ham:SIBCSU}
\end{equation}
where $H_{\text{SI}}$ describes the hopping terms with spin-orbital coupling (SOC)
of the sodium-iridate (SI) type \cite{Shitade2009:prl,Ruegg2012:prl}, $H_{\text{pair}}$ describes the BCS pairing terms, and $H_{\text{int}}$ contains the on-site Hubbard interactions. Explicitly, we have
\begin{subequations}\begin{align}
H_{\text{SI}} = & - t \sum_{\langle ij \rangle s}
c^{\dagger}_{i s} c_{j s} + i \frac{\lambda}{\sqrt{3}}
\sum_{\langle\langle ij \rangle\rangle_{\alpha} s s'}
v_{ij} c^{\dagger}_{is} \sigma^{\alpha}_{ss'} c_{js'} \;,\\
H_{\text{pair}} = & - \Delta_{1} \sum_{\langle ij \rangle s}
c^{\dagger}_{i s} c^{\dagger}_{j s} + i \frac{\Delta_{2}}{\sqrt{3}}
\sum_{\langle\langle ij \rangle\rangle_{\alpha} s s'}
v_{ij} c^{\dagger}_{i s} \sigma^{\alpha}_{ss'} c^{\dagger}_{j s'} + \mathrm{H.c.} \;,\\
H_{\text{int}} = & U \sum_{j} \left( c^{\dagger}_{j\uparrow}c_{j\uparrow} - \frac{1}{2} \right)
\left( c^{\dagger}_{j\downarrow}c_{j\downarrow} - \frac{1}{2} \right) \;,
\end{align}\end{subequations}
where $c^{\dagger}_{j s}$ and $c_{j s}$ are electron creation and annihilation operators at site $j$ with spin polarization $s$, and
$\langle ij \rangle$ and $\langle\langle ij \rangle\rangle$ represent the nearest and next-nearest-neighbor sites respectively.
The first term in $H_{\text{SI}}$ is the usual nearest-neighbor hopping with kinetic energy, and the second term
is the anisotropic SOC, which involves three different Pauli matrices $\sigma^{\alpha}~(\alpha=x,y,z)$ for three inequivalent
next-nearest-neighbor hopping (NNNH) directions. See Figure \ref{fig:sketch}. The sign $v_{ij}=+1$ if the NNNH is anticlockwise with respect to the positive direction of $z$ axis, and $v_{ij}=-1$ if clockwise. $\lambda$ is the SOC strength.
For the superconducting pairing terms, we follow the convention in Ref. \cite{Ezawa2018:prb}. $\Delta_{1,2}$ are pairing
gaps, and $U$ is the on-site Hubbard interaction strength.

We make some remarks on the realization of the model Hamiltonian. The prototype is the SI, a $5d$ transition metal oxide proposed
as a layered correlated QSH insulator \cite{Shitade2009:prl}. When the electron-electron interaction is not too strong,
the SI becomes an interacting topological insulator \cite{Ruegg2012:prl,rachel2018:rpp}. Recent study shows that spin-triplet
$f$-wave pairing can be induced when certain doping is made \cite{kimchi2015:phd}. On the other hand, due to the superconducting proximity, artificial
topological superconductivity can be realized in superconductor-topological insulator heterostructure \cite{FK2008:prl,Xu2014:prl,Sau2010:prb,Stanescu2010:prb,Lababidi2011:prb,Chen2016:prb}.
Therefore, we can consider a bilayer van der Waals heterostructure with the upper layer being the undoped SI with Hubbard interaction, and the lower layer
being the doped topological superconductor. This will induce the BCS pairing term we want as a result of superconducting proximity.
Then we could arrive at an effective Hamiltonian as shown in \cref{ham:SIBCSU} for the upper layer.

Let us now briefly study the symmetry of the model Hamiltonian. Particle hole transformation $\mathcal{C}$ is a unitary transformation
that recombines creation and annihilation operators of fermion, and the operators transforms as
$\mathcal{C} c_{js}^{\dagger} \mathcal{C}^{-1} = (-1)^j c_{js},\mathcal{C} c_{js} \mathcal{C}^{-1} = (-1)^j c_{js}^{\dagger}$,
where the sign $(-1)^j$ is $+1(-1)$ for sublattice $A(B)$ of site $j$ \cite{Chiu2016:rmp,Ryu2008:prb}. One can easily check
that \cref{ham:SIBCSU} is invariant under such a transformation, thus it has particle hole symmetry (PHS). Time reversal
transformation $\mathcal{T}$ is another discrete operation acting on the fermion operators. For spinful systems it reads
$\mathcal{T} = i \sigma^{y} \mathcal{K}$, where $\mathcal{K}$ denotes the complex conjugation, and $\mathcal{T}^2=-1$. It
is known that the SI model Hamiltonian, as a bond-dependent generalization of Kane-Mele (KM) model, is a time reversal invariant
$\mathbb{Z}_{2}$ topological insulator and belongs to the same universal class with the KM model \cite{Shitade2009:prl}.
When the pairing term is taken into consideration, the time reversal symmetry is unaffected. And we will show that the interaction
will not break the time reversal symmetry in \cref{sec:ESU}. Then such a system falls into class DIII in the classification of
topological quantum matter \cite{Chiu2016:rmp}.

\section{Non-Interacting Limit: SI-BCS Model}
\label{sec:SIBCS}

First we investigate the non-interacting limit with $U=0$.
In this case, the model can be called SI-BCS model. For convenience, we will use the following notations. The basis vectors are
\be
\mathbf{a}_{1} =\left(\frac{1}{2},\frac{\sqrt{3}}{2}\right)a,\quad
\mathbf{a}_{2} = \left(-\frac{1}{2},\frac{\sqrt{3}}{2}\right)a,
\ee
where $a$ is the lattice constant and will be set as a unit, $a=1$.
We label the vectors connecting the next-nearest neighbors as $\mathbf{d}_{1}=\mathbf{a}_{1}$,$\mathbf{d}_{2}=-\mathbf{a}_{2}$,
and $\mathbf{d}_{3}=\mathbf{a}_{2}-\mathbf{a}_{1}$, we have
\be
\mathbf{d}_{1} = \left(\frac12, \frac{\sqrt{3}}{2},\right),\;
\mathbf{d}_{2} = \left(\frac12, -\frac{\sqrt{3}}{2},\right),\;
\mathbf{d}_{3} = \left(-1,0\right).\nonumber
\ee
On the other hand, the vectors along the bonds are
\be
\mathbf{e}_{1} = \left(-\frac12, \frac{1}{2\sqrt{3}}\right),\;
\mathbf{e}_{2} = \left(\frac12, \frac{1}{2\sqrt{3}}\right),\;
\mathbf{e}_{3} = \left(0,-\frac{1}{\sqrt{3}}\right).\nonumber
\ee
These two sets of vectors are related by
\be
\mathbf{d}_{1} = \mathbf{e}_{2} - \mathbf{e}_{3},\quad
\mathbf{d}_{2} = \mathbf{e}_{3} - \mathbf{e}_{1},\quad
\mathbf{d}_{3} = \mathbf{e}_{1} - \mathbf{e}_{2}.
\ee
The unit cells are located at $\mathbf{n}=i\mathbf{a}_{1}+j\mathbf{a}_{2}$. Adopting the periodic-boundary condition (PBC), we can define the Fourier transformation as
\begin{equation}
c_{\mathbf{n}} = \frac{1}{\sqrt{L_{x}L_{y}}} \sum_{\mathbf{k}}
c_{\mathbf{k}} \exp(i\mathbf{k}\cdot\mathbf{n}),
\end{equation}
where the wave vector $\mathbf{k}=(k_x,k_y)$, so as to obtain an equivalent Hamiltonian in momentum space. Here we would like to
point out that the SI model has both spin and sublattice degrees of freedom. In the SI-BCS model, the paring terms require us to put the particle and hole creation operators together to form a Nambu spinor as follows
\be
&&\Psi_{\vk}=(\psi_{\vk},\,\psi^{\dag}_{\vk})^T\\
&&\psi_{\vk}=(c_{\vk,A\uparrow},c_{\vk,A\downarrow},c_{\vk,B\uparrow},c_{\vk,B\downarrow})^T
\ee
Therefore the final Bogoliubov-de Gennes Hamiltonian of SI-BCS model is a $8\times8$ matrix which can be written as
\be
H=\left(
    \begin{array}{cccc}
      \lambda G_is_i & -tF s_0 & -\Delta_2 G_is_i & -\Delta_1 F^* s_0 \\
      -tF^* & -\lambda G_is_i & \Delta_1 F s_0  & -\Delta_2 G_is_i \\
      -\Delta_2 G_is_i & \Delta_1 F^*  & \lambda G_is_i & tF \\
      -\Delta_1 F s_0 & -\Delta_2 G_is_i & tF^* s_0 & -\lambda G_is_i
    \end{array}
  \right)
\ee
where the repeated indices means summation over $i=1,2,3$ and $s_i$ are Pauli matrices applying to the spin space and $s^0$ are $2\times2$ identity matrix. Here we have also introduced the following abbreviations,
\be
&&F =\sum_{j=1}^{3} \exp\left( i \mathbf{k}\cdot\mathbf{e}_{j}\right)
=e^{-ik_y/\sqrt{3}}+2e^{ik_y/(2\sqrt{3})}\cos\frac{k_x}2 ,\\
&&G_1 =\frac{2}{\sqrt{3}} \sin{\mathbf{k}\cdot\mathbf{d}_{1}}=\sin\Big(\frac{k_x}2+\frac{\sqrt{3}k_y}2\Big),\\
&&G_2 =\frac{2}{\sqrt{3}}\sin{\mathbf{k}\cdot\mathbf{d}_{2}}=\sin\Big(\frac{k_x}2-\frac{\sqrt{3}k_y}2\Big),\\
&&G_3=\frac{2}{\sqrt{3}}\sin{\mathbf{k}\cdot\mathbf{d}_{3}}=-\sin k_x.
\ee
It is convenient to expand the above Hamiltonian in terms of the spinor basis. To this end, we define the following spinor basis
\be
\Gamma_{abc}=\tau_{a}\otimes\sigma_{b}\otimes s_{c}
\ee
Here the indices $a,b,c=0,\cdots,3$. For $i=1,2,3$ $\tau_i$ and $\sigma_i$ are also Pauli matrices applying to the Nambu pseudo-spin and sublattice space, respectively. $\tau_0$ and $\sigma_0$ are a 2 by 2 identity matrix. Then the Hamiltonian can abbreviated as
\be
H&=&-t\mbox{Re}(F)\Gamma_{310}-t\mbox{Im}(F)\Gamma_{320}\nonumber\\
& &+\Delta_1\mbox{Re}(F)\Gamma_{220}-\Delta_1\mbox{Im}(F)\Gamma_{210}\nonumber\\
& &+\sum_{i=1}^3\lambda G_i\Gamma_{03i}-\sum_{i=1}^3\Delta_{2}G_i \Gamma_{10i}
\label{ham:bloch}
\ee
in the last line the repeated indices are summed, and we find the energy eigenvalues are
\be
E(\vk)=\pm \sqrt{\left(t\pm\Delta_{1}\right)^{2}|F|^2+
\left(\lambda\pm\Delta_{2}\right)^{2}\sum_iG^2_i}
\ee
The band gap closes at two inequivalent Dirac points $K$ and $K'$. In our basis choice, they are given by $K=\left(4\pi/3,0\right)$, $\left(-2\pi/3,\pm 2\pi/\sqrt{3}\right)$ and $K'=\left(-4\pi/3,0\right)$, $\left(2\pi/3,\pm 2\pi/\sqrt{3}\right)$.
We notice that the perfect \textit{flat-band condition} is satisfied when $t=\Delta_{1}$ and $\lambda=\Delta_{2}$.
Later on, we will show that the perfect flat bands condition makes one half of the Majorana fermions decouple from the rest of the Hamiltonian and become localized in the bulk. This in turn leads to an exact solution even if the Hubbard interaction is also included.

\begin{figure}
\centering
\includegraphics[width=\columnwidth]{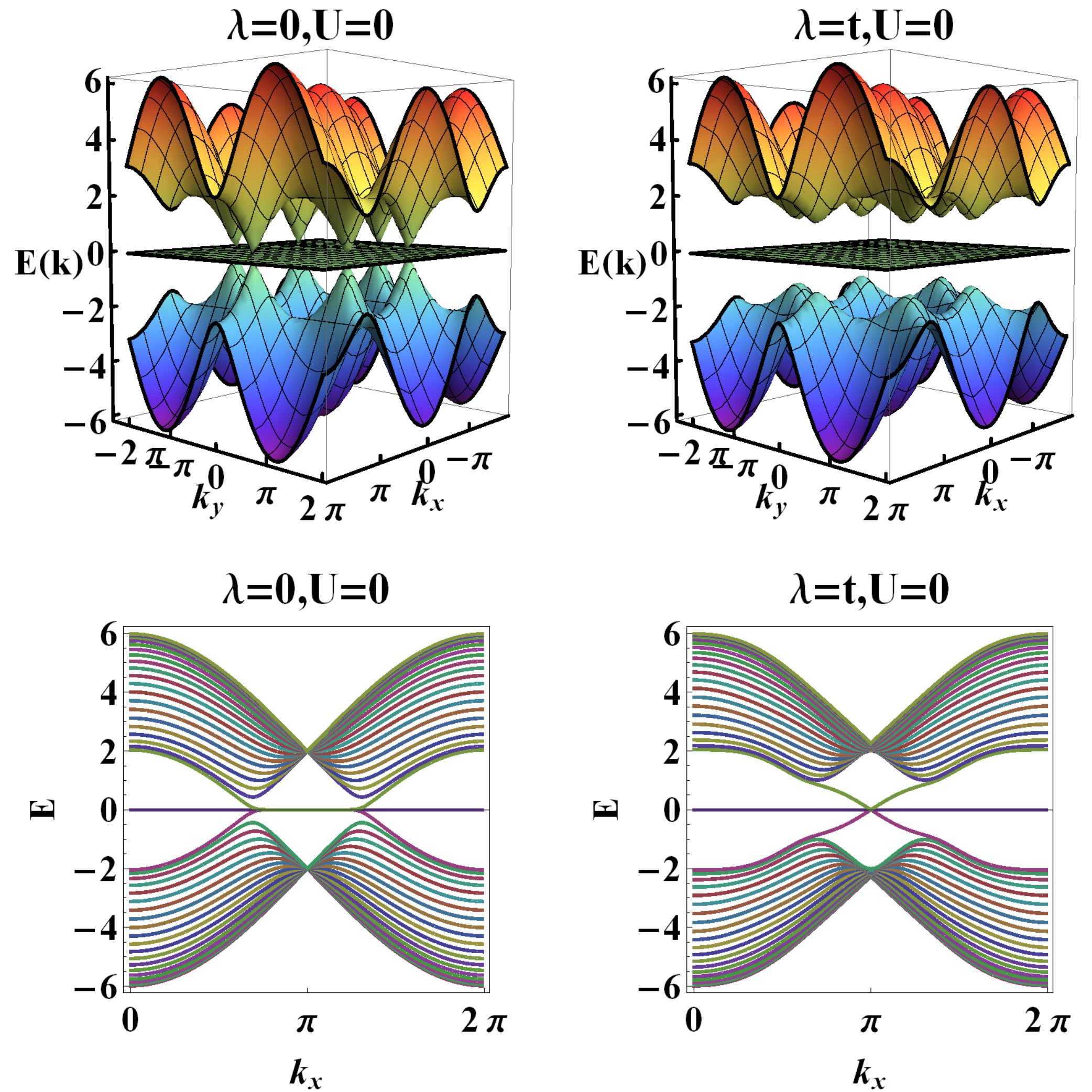}
\caption{Upper panel: Energy eigenvalues in momentum space for $U=0$, $\lambda=0$ (left) and $\lambda=t$ (right).
Lower panels: Energy eigenvalues as a function of $k_x$ with open boundary in $y$ direction for $U=0$, $\lambda=0$ (left) and $\lambda=t$ (right).}
\label{fig:graphenebcs}
\end{figure}

In the following we mainly focus on the system with perfect flat band. If the SOC is turned off, the system reduces to a Dirac-nodal superconductor, which can be described by the Graphene-BCS model, and the energy gap closed linearly at the $K$ and $K'$ point. The energy spectrum is obtained in both torus and cylinder geometry, i.e., PBC is imposed in both $x$ and $y$ direction, or with open-boundary condition (OBC) in $y$ direction, as illustrated in \cref{fig:graphenebcs}.

The upper panels shows the dispersions in the momentum space, while the lower panels displays the energy spectrum as a function of $k_x$ with open boundary in $y$ direction. In these momentum space plots, there are always four topologically trivial zero-energy flat bands lying in the middle. In the left two panels, the SOC and Hubbard interaction strength are all set to be zero. Thus, we obtain a graphene-BCS model. In the upper left panel, the two-fold degenerate Graphene-like conduction bands and valence bands linearly touch with each other at $K$ and $K'$ points in the Brillouin zone. Correspondingly, the lower left plots with open boundary in $y$ direction shows that the conduction band and valence band touch on the line connecting the two inequivalent Dirac points. Therefore it describes a nodal superconductor.

If the SOC is introduced, the Dirac fermions at $K$ and $K'$ point obtain a mass, meanwhile an energy gap opens up. In this case the model is described by SI-BCS Hamiltonian. In the upper right panel, one can see that the two-fold degenerate conduction bands and valence bands are fully gapped in the bulk. In the lower right panel with open boundary in $y$ direction, there are gapless helical edge states connecting the valence and conduction bands, which indicates non-trivial topological property of the model.

\begin{figure}
\centering
\includegraphics[width=5cm]{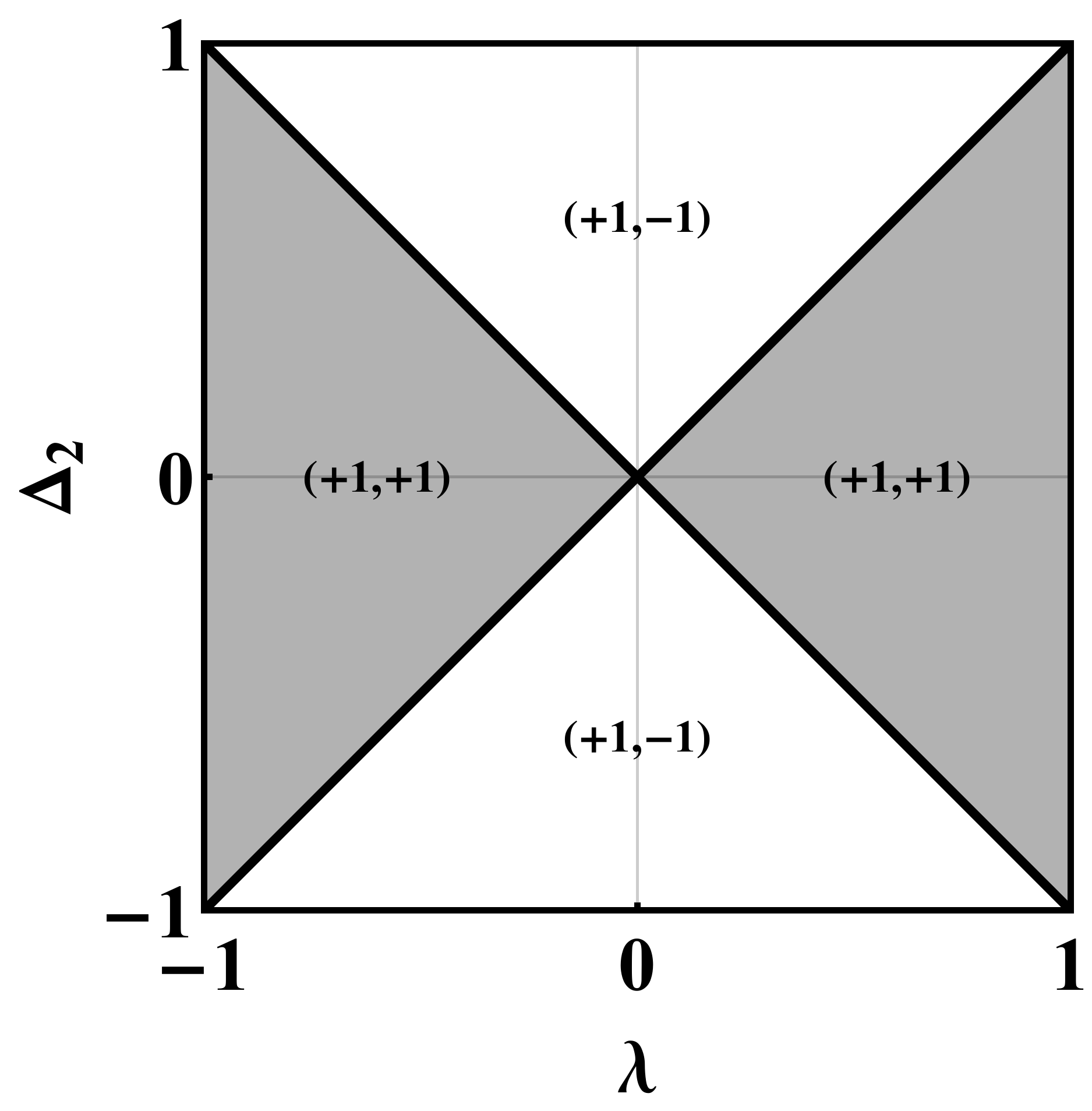}
\caption{Phase diagram. Different phases are labeled by isospin Chern number $(C_{+,\sigma},C_{-,\sigma})$.
On the two critical lines, one of the isospin Chern number becomes zero. The origin is a multi-critical point.}
\label{fig:phase}
\end{figure}

We now investigate the topological properties of the system. The topological invariant describing class DIII
topological superconductors is the $\mathbb{Z}_2$ index \cite{Chiu2016:rmp,Ryu2008:prb,Qi2009:prl},
or spin Chern number equivalently \cite{Sheng2005:prl,Sheng2006:prl,Prodan2009:prb}.
Making the Taylor expansion near the $K$ point, one gets the effective Hamiltonian
\begin{equation}\bea
\mathcal{H}_{\mathbf{k}} = &-\eta v_F k_{x} \Gamma_{310}-v_F k_{y} \Gamma_{320} \\
& + \eta \tilde{\Delta}_{1} k_{x} \Gamma_{220}-\tilde{\Delta}_1 k_{y} \Gamma_{210} \\
& + \lambda \eta \sum_{j} \Gamma_{03j}-\Delta_{2}\eta \sum_{j} \Gamma_{10j},
\label{ham:effHam}
\eea\end{equation}
Here $\eta=\pm1$ for $K$ and $K'$ point respectively. For convenience, we define $v_F=\sqrt{3}t/2$ and $\tilde{\Delta}_1=\sqrt{3}\Delta_1/2$.
Then we can make the unitary transformation $\widetilde{\mathcal{H}}_{s}(\mathbf{k}) =
\mathcal{U}^{-1} \mathcal{H}_{\mathbf{k}} \mathcal{U}$
with $\mathcal{U}=\exp(i \frac{\pi}{8} \Gamma_{002}) \exp (i \frac{\pi}{8} \Gamma_{003})$.
The resulting $\widetilde{\mathcal{H}}_{s}(\mathbf{k})$ is a block diagonal matrix in the form,
\begin{equation}
\widetilde{\mathcal{H}}_{s}(\mathbf{k})=
\left(\begin{array}{cccc}
s\eta\lambda & v_F k_{\eta} & -s\eta\Delta_{2} & \tilde{\Delta}_1 k_{\eta}\\
v_F k_{\eta}^{\ast} & -s\eta\lambda & -\tilde{\Delta}_1 k_{\eta}^{\ast} & -s\eta\Delta_{2}\\
-s\eta\Delta_{2} & -\tilde{\Delta}_1 k_{\eta} & s\eta\lambda & -v_F k_{\eta}\\
\tilde{\Delta}_1 k_{\eta}^{\ast} & -s\eta\Delta_{2} & -v_F k_{\eta}^{\ast} & -s\eta\lambda
\end{array}\right)
\end{equation}
Here $k_{\eta}=\eta k_{x}-ik_{y}$.
Now the energy eigenvalues for the bands with sign $\pm$ are
$$E_{\pm}(\mathbf{k})=\sqrt{\left(1\pm\Delta_{1}/t\right)^{2}v^{2}_{F}k^{2}+
\left(\lambda\pm\Delta_{2}\right)^{2}}.$$
The eigen-wavefunctions for the two valence bands are
\begin{equation}
\left\vert \psi_{\pm}(\mathbf{k})\right\rangle
=\left[\begin{array}{c}
\sin\alpha_{\pm}(k)\\
-\cos\alpha_{\pm}(k)e^{i\eta\theta(k)}\\
\sin\alpha_{\pm}(k)\\
\cos\alpha_{\pm}(k)e^{i\eta\theta(k)}
\end{array}\right],
\end{equation}
where $2\alpha_{\pm}(k)=\arctan\frac{ \sqrt{3} \left(t\pm\Delta_{1}\right)k/2}{s\eta\left(\lambda\pm\Delta_{2}\right)}$
and $e^{i\eta\theta(k)}=\left(\eta k_{x}-ik_{y}\right)/k$.
The Berry connection is defined as $\mathcal{A}_{\pm}(\mathbf{k})=-i\left\langle \psi_{\pm}(\mathbf{k})\right|\nabla_{\mathbf{k}}\left\vert \psi_{\pm}(\mathbf{k})\right\rangle$
and the Berry curvature is
\begin{equation}\bea
\mathcal{F}_{\pm}(\mathbf{k})=&\nabla_{\mathbf{k}}\times\mathcal{A}_{\pm}(\mathbf{k}) \\
=&\frac{s\left(\lambda\pm\Delta_{2}\right)(1\pm\Delta_{1}/t)v_F^2}
{2\left[\left(1\pm\Delta_{1}/t\right)^{2}v^{2}_{F}k^{2}
+\left(\lambda\pm\Delta_{2}\right)^{2}\right]^{3/2}}
\eea\end{equation}
Then the Chern number is the integral over the polar plane
\begin{align}
C_{\pm,s}=\int\mathcal{F}_{\pm}(\mathbf{k})\mathrm{d}^2 \mathbf{k}/2\pi=\frac{1}{2}\operatorname{sgn}\left[s\left(\lambda\pm\Delta_{2}\right)\right]
\end{align}
with $\operatorname{sgn}(x)=\lim_{\varepsilon\to0}x/\sqrt{x^2+\varepsilon^2}$ bring the sign function.
The total Chern number and isospin Chern number for each band is defined as
\begin{subequations}\begin{align}
C_{\pm} = & C_{\pm,s}+C_{\pm,\bar{s}} \\
C_{\pm,\sigma} = & C_{\pm,s}-C_{\pm,\bar{s}}
\end{align}\end{subequations}

The phase diagram is shown in \cref{fig:phase}. As a result of time reversal symmetry, the total Chern number is always zero.
If $|\lambda|>|\Delta_2|$, the spin Chern number is 1 for both two bands, which indicates the system is a helical topological superconductor.
While if $|\lambda|<|\Delta_2|$, the spin Chern number has opposite spin Chern number, and the system turn out to be.
And we see that if the perfect flat band condition is satisfied, the Chern number of the flat band is zero, and system is always topological
with nonzero $\lambda$.

\section{Exact Solution of Interacting Model Along Symmetric Lines}
\label{sec:ESU}

\begin{figure}
\centering
\includegraphics[width=8cm]{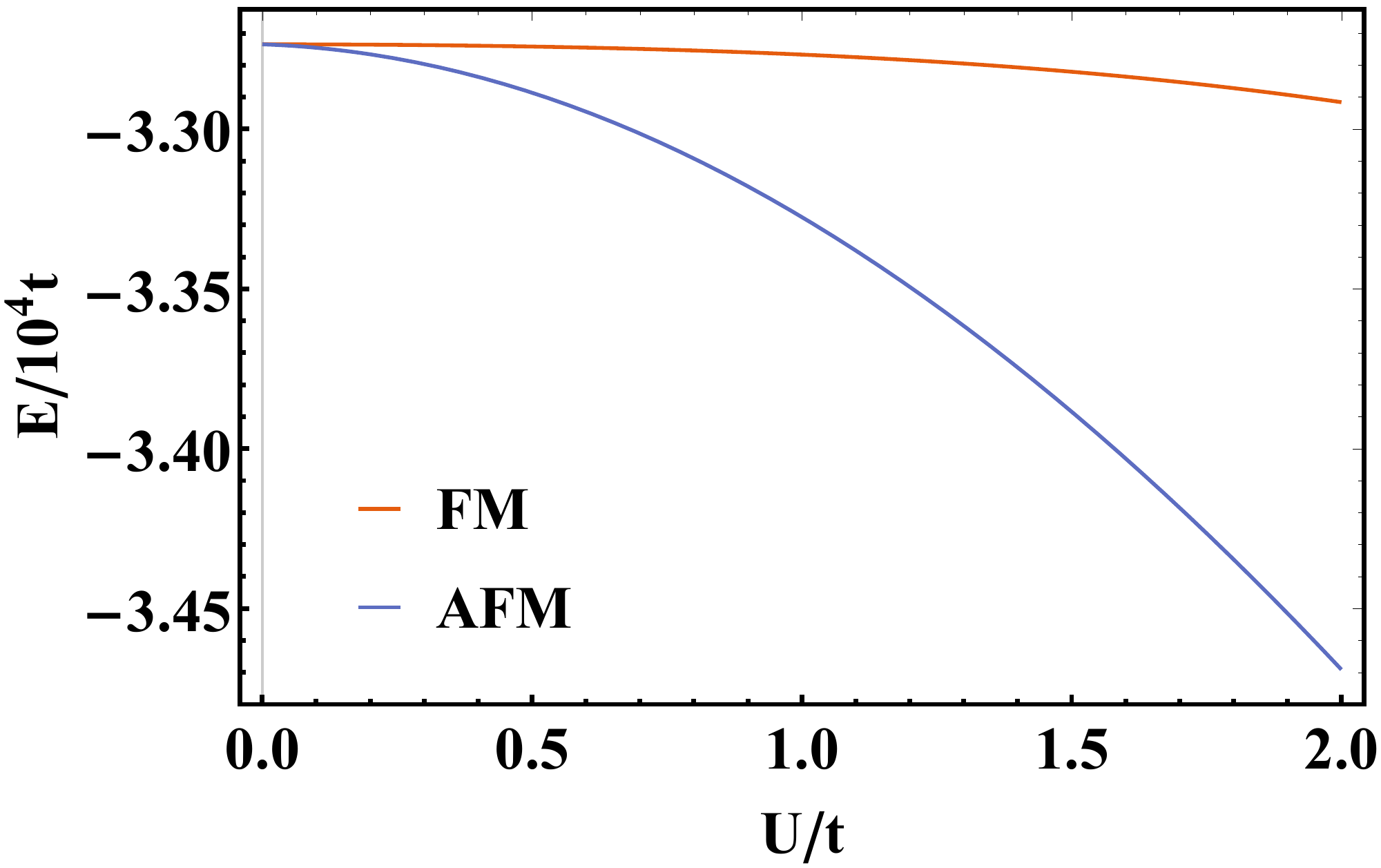}
\caption{Comparison of ground state energy with AFM and FM order. We chose $\lambda=0.4t$.}
\label{fig:gs}
\end{figure}

\begin{figure}
\centering
\includegraphics[width=\columnwidth]{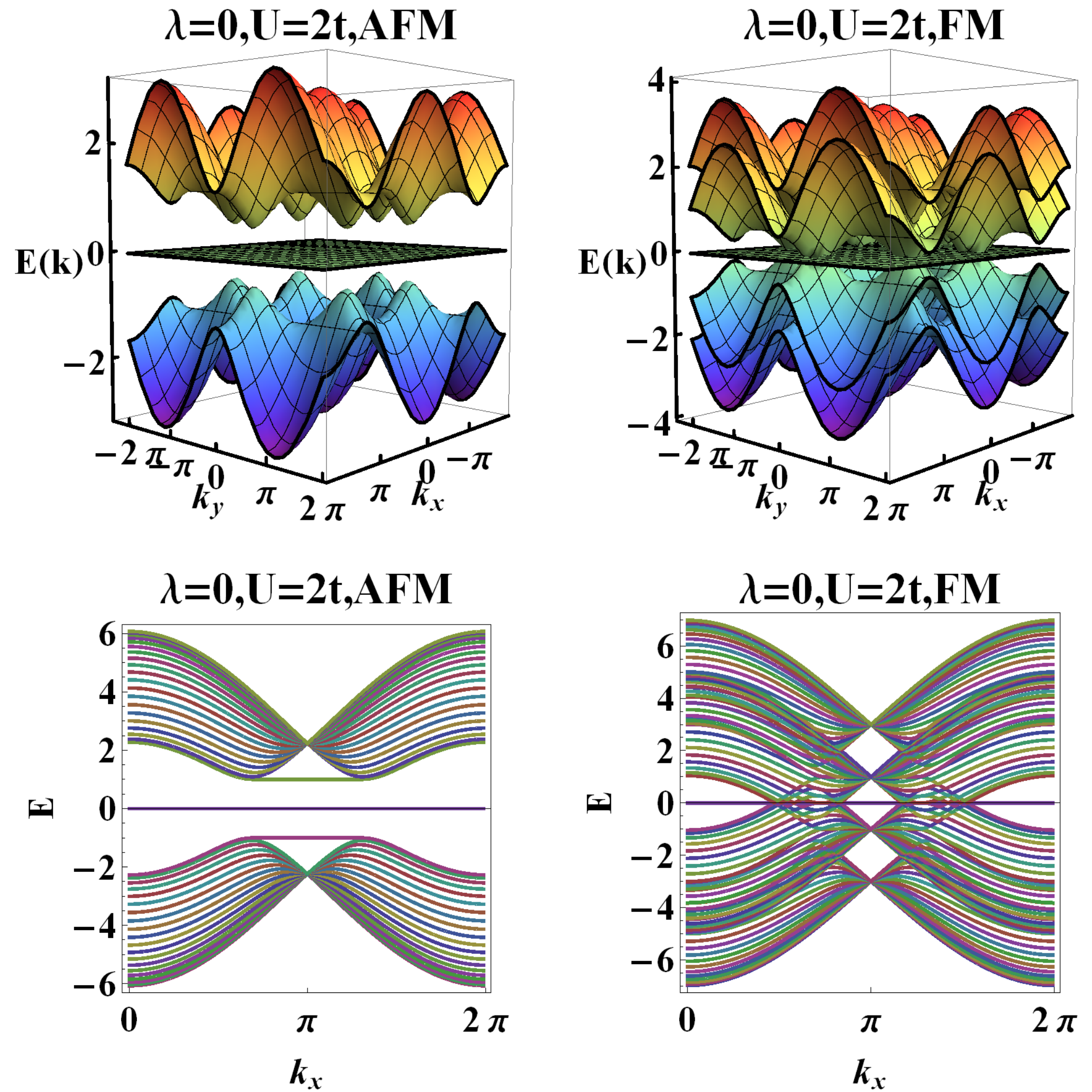}
\caption{Band structure of the Graphene-BCS-Hubbard model with AFM or FM order.
The upper panel are the energy eigenvalues in momentum space and below are
the corresponding spectra in nanoribbon geometry. The ground state is a full gapped
superconductor. And the lack of crossing edge states indicates topological triviality.}
\label{fig:graphenebcsU}
\end{figure}

\begin{figure}
\centering
\includegraphics[width=\columnwidth]{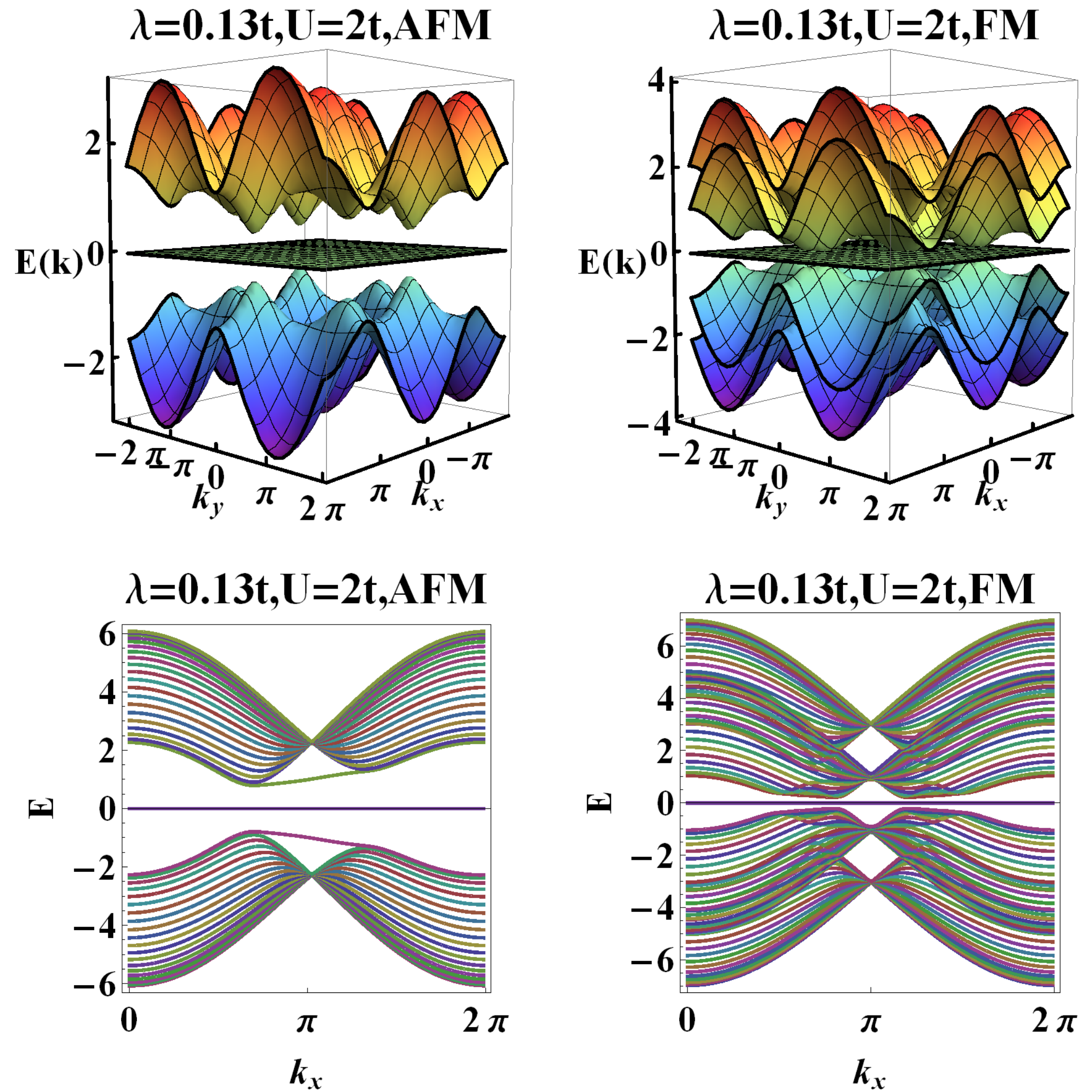}
\caption{Band structure of SI-BCS-Hubbard model with AFM or FM order.
The upper panel are the energy eigenvalues in momentum space and below are
the corresponding spectra in nanoribbon geometry.}
\label{fig:SIbcsU}
\end{figure}

In this section we show that the model above is exactly solvable even
with the Hubbard interaction being taken into account.
By introducing the Majorana fermion operators $\eta$ and $\gamma$ for each sublattice site,
\begin{subequations}\begin{align}
c_{\mathbf{n},A\sigma} = & \eta_{\mathbf{n},A\sigma} + i \gamma_{\mathbf{n},A\sigma} \;,\;
c^{\dagger}_{\mathbf{n},A\sigma} = \eta_{\mathbf{n},A\sigma} - i \gamma_{\mathbf{n},A\sigma} \;,\;\\
c_{\mathbf{n},B\sigma} = & \gamma_{\mathbf{n},B\sigma} + i \eta_{\mathbf{n},B\sigma} \;,\;
c^{\dagger}_{\mathbf{n},B\sigma} = \gamma_{\mathbf{n},B\sigma} - i \eta_{\mathbf{n},B\sigma} \;,\;
\end{align}\end{subequations}
the Hamiltonian in \cref{ham:SIBCSU} can be rewritten as
\begin{equation}
H = H_{1} + H_{2} + H_{3}
\end{equation}
with
\begin{subequations}\begin{align}
H_{1} = & 2i\sum_{\langle ij \rangle s}
\left(\Delta_{1}+t\right) \gamma_{i s}\gamma_{j s} +
\left(\Delta_{1}-t\right) \eta_{i s}\eta_{j s}, \\
H_{2} = & \frac{i}{2\sqrt{3}} \sum_{\langle\langle ij
\rangle\rangle_{\alpha} s s'}
v_{ij} \left(\lambda+\Delta_{2}\right)
\gamma_{i s}\sigma^{\alpha}_{ss'}\gamma_{j s'} + \nonumber \\
& v_{ij} \left(\lambda-\Delta_{2}\right)
\eta_{i s}\sigma^{\alpha}_{ss'}\eta_{j s'}, \\
H_{3} = & U\sum_{i} \left(2i\eta_{i\uparrow}\gamma_{i\uparrow}\right)
\left(2i\eta_{i\downarrow}\gamma_{i\downarrow}\right).
\end{align}\end{subequations}
Notice that the commutation relations are
\begin{equation}
\left\{\eta_{\mathbf{m},A\alpha},\eta_{\mathbf{n},B\beta}\right\} =
\left\{\gamma_{\mathbf{m},A\alpha},\gamma_{\mathbf{n},B\beta}\right\} =
\frac{1}{2}\delta_{\mathbf{m}\mathbf{n}} \delta_{AB} \delta_{\alpha\beta}.
\label{eq:cc}
\end{equation}
By requiring the flat-band condition, $t=\Delta_{1}$ and $\lambda=\Delta_{2}$, we immediately find that the $\eta$
Majorana fermions disappear in $H_{1}$ and $H_{2}$. To make it more clear, we can introduce
$D_{j}=4i\eta_{j\uparrow}\eta_{j\downarrow}$ since it commutes with $H_{3}$ for all sites $j$, i.e. it becomes a $c$-number. Then with the commutation relation in \cref{eq:cc} being taken into consideration,
we find that $D^{2}_{j}=1$ and hence $D_{j}=\pm 1$. Finally, we arrive at the total Hamiltonian,
\begin{equation}\bea
H = & 4it \sum_{\langle ij \rangle} \sum_{\sigma} \gamma_{i\sigma}\gamma_{j\sigma} +
\frac{i\lambda}{\sqrt{3}} \sum_{\langle\langle ij \rangle\rangle_{\alpha} s s'}
v_{ij} \gamma_{i\alpha}\sigma^{\gamma}_{\alpha\beta}\gamma_{j\beta} \\
& -iU \sum_{j} D_{j} \left(\gamma_{j\uparrow}\gamma_{j\downarrow}\right)
\label{ham:fbham}
\eea\end{equation}
The decouple of $\eta$ Majorana fermions makes the original interaction terms to be quadratic
and thus \cref{ham:fbham} is exactly solvable for any fixed set of $D_{j}$,
which serves as a background $\mathbb{Z}_{2}$ gauge field. For a $N$-site system there
are $2^N$ choices of the set $D_{j}$, and the total Hilbert space is the direct product
of that of one Hamiltonian with certain $D_{j}$'s configuration. Two uniform configurations
are of most interest. One is the ferromagnetic (FM) configuration with uniform $D_j=1$ for both of the two
sublattices $A$ and $B$. Another is the antiferromagnetic (AFM) configuration, in which the signs of $D_j$'s are
opposite for the two sublattices. In the basis of Nambu spinor, $H_{3}$ can be written as
\begin{equation}
H_{\textrm{FM}} = - \frac{U}{4} \left(\Gamma_{002}-\Gamma_{132}\right) \;,\;
H_{\textrm{AFM}} = - \frac{U}{4} \left(\Gamma_{032}-\Gamma_{102}\right)
\label{ham:fmafm}
\end{equation}
for the two particular orders. And the eigenvalues are
\begin{subequations}\begin{align}
E^{2}_{\text{FM},\pm} = & t^2 |F|^2 + \lambda^2\sum_{j=1}^3G_j^2 +
\left(\frac{U}{4}\right)^2  \nonumber \\ & \pm \frac{U}{2} \sqrt{t^2 |F|^2 +
\lambda^2 G_2^2} \\
E^{2}_{\text{AFM}} = & t^2 |F|^2 + \lambda^2\sum_{j=1}^3G_j^2 +
\left(\frac{U}{4}\right)^2 + \frac{\lambda U}{2}G_2
\end{align}\end{subequations}
We see that for ferromagnetic order, the background $D_j$ splits the energy
bands for different isospin, while for the anti-ferromagnetic order the bands
remains to be two-fold degenerate. By using the arithmetic mean-root mean square inequality
$$
\sum_{\pm} E_{\text{FM},\pm} \leqslant \sqrt{2\sum_{\pm} E^{2}_{\text{FM},\pm}}
< 2 E_{\text{AFM}}
$$
we explicitly show that the AFM order has lower ground state energy
than that of FM order, then the ground state fall into the Hilbert
space of the Hamiltonians with AFM configuration. Such a configuration
does not satisfy the time reversal symmetry, so the time reversal symmetry
is spontaneously broken.

We now study the band structure of the interacting topological superconductors with time reversal symmetry being broken. When the SOC strength $\lambda=0$, the model
reduces to the Graphene-BCS-Hubbard model. The energy band is shown in \cref{fig:graphenebcsU}. For the AFM case, the interactions can be treated as a staggered potential on isospins and thus the Graphene-like partial flat bands are moved away from zero energy but the isospin degeneracy is not affected. Such a system is a
full gapped superconductor. While for the FM case, the isospin degeneracy is lifted, meanwhile the conduction and valence bands touch with each other along a circle around the $K$ points. Such a system is a loop-nodal superconductor. Because of the absence of SOC, both above mentioned two cases are topologically trivial.

Then we moved on to investigate the cases with SOC being turned on. The results are shown in Figure \ref{fig:SIbcsU}. In the AFM case, the presence of SOC makes the spin-up and spin-down configurations in $y$ direction inequivalent. And because of this, the Hubbard interaction plays a role of gate voltage in $y$ direction, which opens up a band gap and makes the two $K$ valleys imbalanced. Now the energy eigenvalues is not symmetric, i.e. $E(k_x)\neq E(-k_x)$, which is the consequence of the time reversal symmetry breaking. While for the FM case the band is symmetric.

Although the time reversal symmetry breaking makes the $\mathbb{Z}_{2}$ index ill-defined, the isospin Chern number may be used to describe the non-trivial topological properties. For the superconductor with AFM order, the band gap will not close and re-open again. This is ascribed to a quantum phase transition into a trivial topological superconductor, just as described in Refs. \cite{rachel2016:jpc,ezawa2013:sr}. On the other hand, topological non-trivial phase could survive in FM order \cite{Sheng2011:prl,Sheng2013:cpb}. To explore whether there are some topological nontrivial phases in the FM order, we first take a close look at the gap-closing condition. The necessary condition for gap-closing is
$$
\sin k_{x} = 0\;,\; \sin\left(\frac{1}{2}k_{x}\pm\frac{\sqrt{3}}{2}k_{y}\right)=0
$$
and we see that the SOC strength is not involved. At the momentum determined by the above equation, the requirement of gap-closing leads to
$t|F|\pm U=0$  Therefore, we find that at two special interaction strength $U=4t$ or $U=12t$, the conduction and valence bands will touch with each other.

\begin{figure}
\centering
\includegraphics[width=8cm]{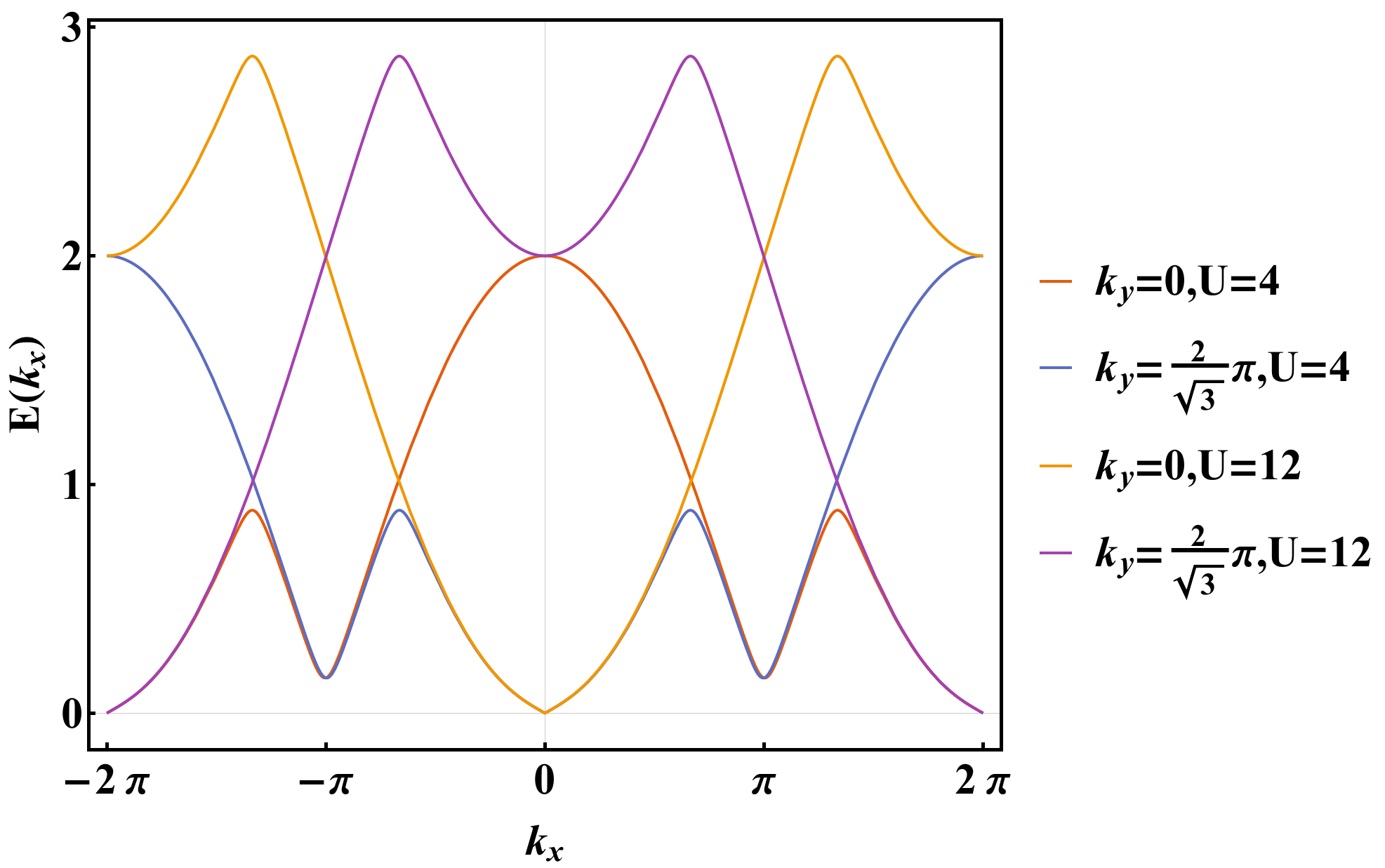}
\caption{The band energy $E$ with FM configuration as a function of $k_x$ for $U=4t$, $U=12t$ and $k_y=0$ and $\frac{2}{\sqrt{3}}\pi$.}
\label{fig:gapclosing}
\end{figure}

\begin{figure}
\centering
\includegraphics[width=8cm]{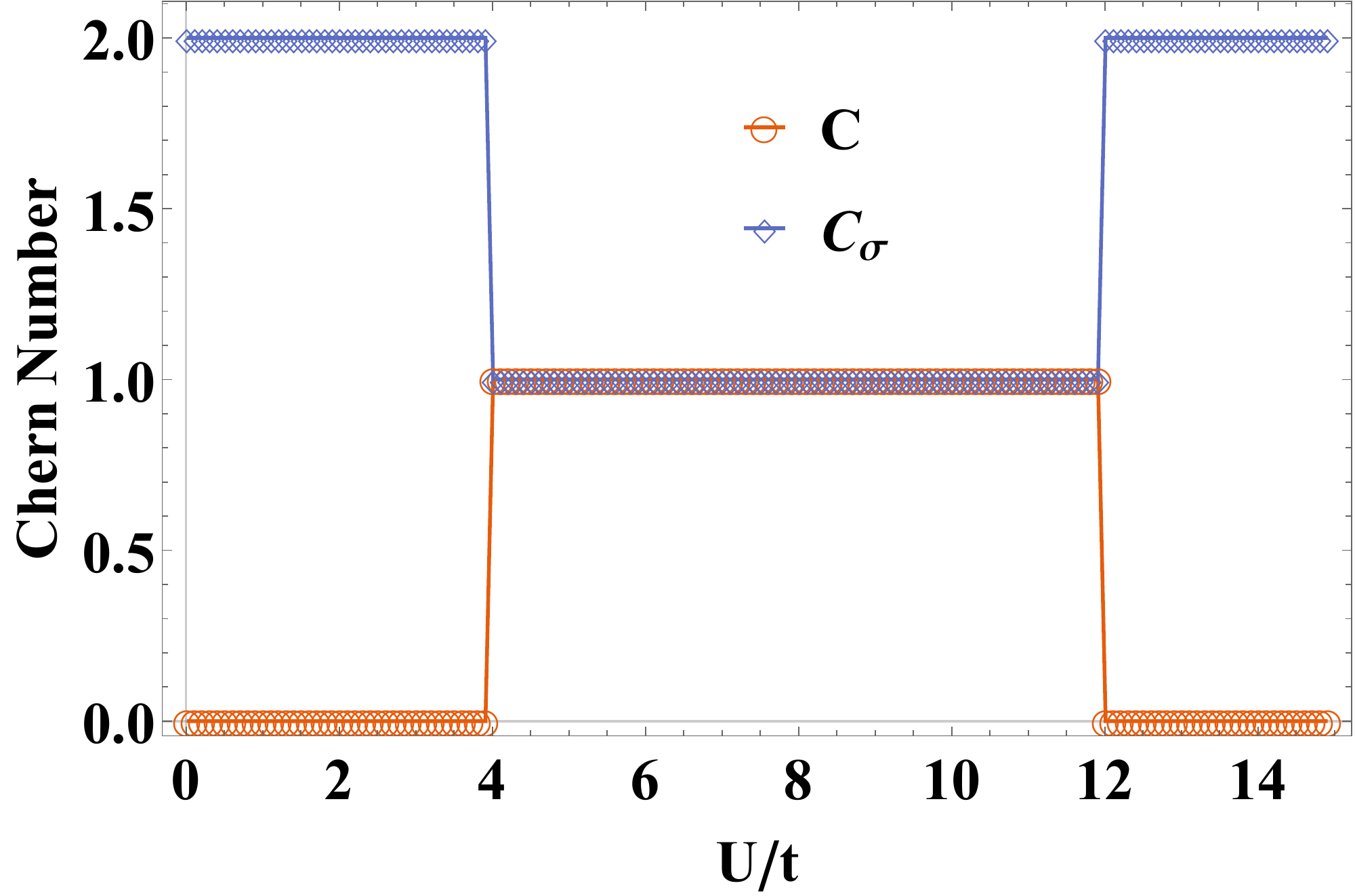}
\caption{The total Chern number and isospin Chern number of SI-BCS-Hubbard model with FM configuration as a function of $U/t$.}
\label{fig:FMtop}
\end{figure}

\begin{figure}
\centering
\includegraphics[width=8cm]{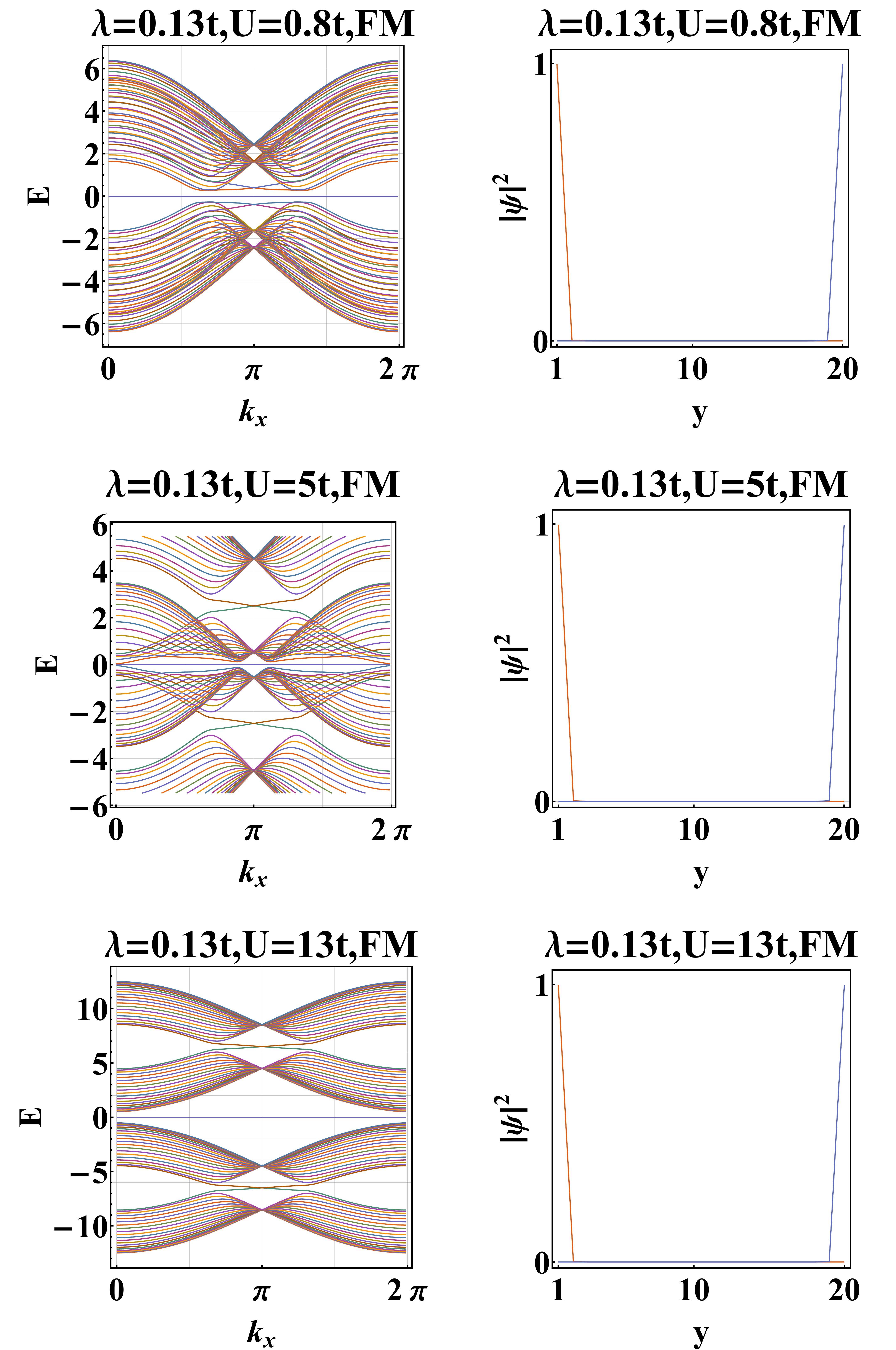}
\caption{The left column: Band structures of SI-BCS-Hubbard model with FM order and cylinder geometry for $U/t=0.8, 5, 13$.
The right column: The amplitudes of wave-function of the edge states around $k_x=\pi$.}
\label{fig:edge}
\end{figure}

To further check that whether the gap-closing could induce a topological transition, we numerically compute the
Chern numbers for the valence bands, and the result is shown in \cref{fig:FMtop}. We see that the total Chern numbers
are zero for $U/t<4$ or $U/t>12$, while the isospin Chern numbers are 2, indicating the quantum spin Hall phase.
Surprisingly, we find that for $4<U/t<12$, the total Chern number becomes 1, indicating the quantum anomalous Hall
phase. To clarify the mystery, we reinvestigate the band structures and localized edge states to see how the interaction
term affects the topological properties. We choose three points $U/t=0.8,5$ and $13$, and the band structures together
with corresponding localized edge states are plotted in \cref{fig:edge}. Both edge states are taken in gap from the valence
bands with momentum $k_{x}=\pi$.

The three representative diagrams help us build a better understanding of the evolution of band structures under the interaction.
Roughly speaking, the FM interaction term in \cref{ham:fmafm} can be regarded as a chemical potential or gate voltage in the spin-$y$
direction. When $U/t<4$ the interaction strength is weak and the degeneracy of two isospins are lifted. The two edge states crossing at $k_{x}=\pi$
connects two inequivalent valley of two bands with different isospins. The system is in a quantum spin Hall phase.
With the interaction strength increasing to $4<U/t<12$, we can see that the bands are strongly inverted.
The electrons become full polarized and results in the quantum anomalous Hall phase.
Finally, when the interaction strength is so strong that the conduction band and valence band with
the same isospin are completed inverted. Therefore, the quantum spin Hall phase is restored.

\section{Conclusion}
\label{sec:con}

In conclusion, we have proposed a SI model on the honeycomb lattice with both BCS pairing potential and Hubbard interaction term to explore the nontrivial topology in an interacting system. Due to the Hubbard interaction, the time reversal symmetry is spontaneous symmetry broken. The ground state of this model is a solution with anti-ferromagnetic configuration. For the solution with ferromagnetic configuration, nontrivial topology is found and characterized by the isospin Chern number.

\acknowledgments

This work is supported by NSFC under Grant No. 11874272.


\bibliography{ref}

\begin{thebibliography}{36}%
\makeatletter
\providecommand \@ifxundefined [1]{%
 \@ifx{#1\undefined}
}%
\providecommand \@ifnum [1]{%
 \ifnum #1\expandafter \@firstoftwo
 \else \expandafter \@secondoftwo
 \fi
}%
\providecommand \@ifx [1]{%
 \ifx #1\expandafter \@firstoftwo
 \else \expandafter \@secondoftwo
 \fi
}%
\providecommand \natexlab [1]{#1}%
\providecommand \enquote  [1]{``#1''}%
\providecommand \bibnamefont  [1]{#1}%
\providecommand \bibfnamefont [1]{#1}%
\providecommand \citenamefont [1]{#1}%
\providecommand \href@noop [0]{\@secondoftwo}%
\providecommand \href [0]{\begingroup \@sanitize@url \@href}%
\providecommand \@href[1]{\@@startlink{#1}\@@href}%
\providecommand \@@href[1]{\endgroup#1\@@endlink}%
\providecommand \@sanitize@url [0]{\catcode `\\12\catcode `\$12\catcode
  `\&12\catcode `\#12\catcode `\^12\catcode `\_12\catcode `\%12\relax}%
\providecommand \@@startlink[1]{}%
\providecommand \@@endlink[0]{}%
\providecommand \url  [0]{\begingroup\@sanitize@url \@url }%
\providecommand \@url [1]{\endgroup\@href {#1}{\urlprefix }}%
\providecommand \urlprefix  [0]{URL }%
\providecommand \Eprint [0]{\href }%
\providecommand \doibase [0]{http://dx.doi.org/}%
\providecommand \selectlanguage [0]{\@gobble}%
\providecommand \bibinfo  [0]{\@secondoftwo}%
\providecommand \bibfield  [0]{\@secondoftwo}%
\providecommand \translation [1]{[#1]}%
\providecommand \BibitemOpen [0]{}%
\providecommand \bibitemStop [0]{}%
\providecommand \bibitemNoStop [0]{.\EOS\space}%
\providecommand \EOS [0]{\spacefactor3000\relax}%
\providecommand \BibitemShut  [1]{\csname bibitem#1\endcsname}%
\let\auto@bib@innerbib\@empty
\bibitem [{\citenamefont {Hasan}\ and\ \citenamefont
  {Kane}(2010)}]{Kane-review}%
  \BibitemOpen
  \bibfield  {author} {\bibinfo {author} {\bibfnamefont {M.~Z.}\ \bibnamefont
  {Hasan}}\ and\ \bibinfo {author} {\bibfnamefont {C.~L.}\ \bibnamefont
  {Kane}},\ }\href@noop {} {\bibfield  {journal} {\bibinfo  {journal} {Rev.
  Mod. Phys.}\ }\textbf {\bibinfo {volume} {82}},\ \bibinfo {pages} {3045}
  (\bibinfo {year} {2010})}\BibitemShut {NoStop}%
\bibitem [{\citenamefont {Qi}\ and\ \citenamefont
  {Zhang}(2011)}]{Zhang-review}%
  \BibitemOpen
  \bibfield  {author} {\bibinfo {author} {\bibfnamefont {X.-L.}\ \bibnamefont
  {Qi}}\ and\ \bibinfo {author} {\bibfnamefont {S.-C.}\ \bibnamefont {Zhang}},\
  }\href@noop {} {\bibfield  {journal} {\bibinfo  {journal} {Rev. Mod. Phys.}\
  }\textbf {\bibinfo {volume} {83}},\ \bibinfo {pages} {1057} (\bibinfo {year}
  {2011})}\BibitemShut {NoStop}%
\bibitem [{\citenamefont {Schnyder}\ \emph
  {et~al.}(2008{\natexlab{a}})\citenamefont {Schnyder}, \citenamefont {Ryu},
  \citenamefont {Furusaki},\ and\ \citenamefont {Ludwig}}]{Ryu2008}%
  \BibitemOpen
  \bibfield  {author} {\bibinfo {author} {\bibfnamefont {A.}~\bibnamefont
  {Schnyder}}, \bibinfo {author} {\bibfnamefont {S.}~\bibnamefont {Ryu}},
  \bibinfo {author} {\bibfnamefont {A.}~\bibnamefont {Furusaki}}, \ and\
  \bibinfo {author} {\bibfnamefont {A.}~\bibnamefont {Ludwig}},\ }\href@noop {}
  {\bibfield  {journal} {\bibinfo  {journal} {Phys. Rev. B}\ }\textbf {\bibinfo
  {volume} {78}},\ \bibinfo {pages} {195125} (\bibinfo {year}
  {2008}{\natexlab{a}})}\BibitemShut {NoStop}%
\bibitem [{\citenamefont {Kitaev}(2009)}]{Kitaev-AIP}%
  \BibitemOpen
  \bibfield  {author} {\bibinfo {author} {\bibfnamefont {A.}~\bibnamefont
  {Kitaev}},\ }\href@noop {} {\bibfield  {journal} {\bibinfo  {journal} {AIP
  conf. Proc.}\ }\textbf {\bibinfo {volume} {22}},\ \bibinfo {pages} {1134}
  (\bibinfo {year} {2009})}\BibitemShut {NoStop}%
\bibitem [{\citenamefont {Chiu}\ \emph {et~al.}(2016)\citenamefont {Chiu},
  \citenamefont {Teo}, \citenamefont {Schnyder},\ and\ \citenamefont
  {Ryu}}]{Chiu2016:rmp}%
  \BibitemOpen
  \bibfield  {author} {\bibinfo {author} {\bibfnamefont {C.-K.}\ \bibnamefont
  {Chiu}}, \bibinfo {author} {\bibfnamefont {J.~C.~Y.}\ \bibnamefont {Teo}},
  \bibinfo {author} {\bibfnamefont {A.~P.}\ \bibnamefont {Schnyder}}, \ and\
  \bibinfo {author} {\bibfnamefont {S.}~\bibnamefont {Ryu}},\ }\href {\doibase
  10.1103/RevModPhys.88.035005} {\bibfield  {journal} {\bibinfo  {journal}
  {Rev. Mod. Phys.}\ }\textbf {\bibinfo {volume} {88}},\ \bibinfo {pages}
  {035005} (\bibinfo {year} {2016})}\BibitemShut {NoStop}%
\bibitem [{\citenamefont {Fidkowski}\ and\ \citenamefont
  {Kitaev}(2010)}]{Kitaev-1}%
  \BibitemOpen
  \bibfield  {author} {\bibinfo {author} {\bibfnamefont {L.}~\bibnamefont
  {Fidkowski}}\ and\ \bibinfo {author} {\bibfnamefont {A.}~\bibnamefont
  {Kitaev}},\ }\href@noop {} {\bibfield  {journal} {\bibinfo  {journal} {Phys.
  Rev. B}\ }\textbf {\bibinfo {volume} {81}},\ \bibinfo {pages} {134509}
  (\bibinfo {year} {2010})}\BibitemShut {NoStop}%
\bibitem [{\citenamefont {Fidkowski}\ and\ \citenamefont
  {Kitaev}(2011)}]{Kitaev-2}%
  \BibitemOpen
  \bibfield  {author} {\bibinfo {author} {\bibfnamefont {L.}~\bibnamefont
  {Fidkowski}}\ and\ \bibinfo {author} {\bibfnamefont {A.}~\bibnamefont
  {Kitaev}},\ }\href@noop {} {\bibfield  {journal} {\bibinfo  {journal} {Phys.
  Rev. B}\ }\textbf {\bibinfo {volume} {83}},\ \bibinfo {pages} {075103}
  (\bibinfo {year} {2011})}\BibitemShut {NoStop}%
\bibitem [{\citenamefont {Yao}\ and\ \citenamefont {Ryu}(2013)}]{Yao}%
  \BibitemOpen
  \bibfield  {author} {\bibinfo {author} {\bibfnamefont {H.}~\bibnamefont
  {Yao}}\ and\ \bibinfo {author} {\bibfnamefont {S.}~\bibnamefont {Ryu}},\
  }\href@noop {} {\bibfield  {journal} {\bibinfo  {journal} {Phys. Rev. B}\
  }\textbf {\bibinfo {volume} {88}},\ \bibinfo {pages} {064507} (\bibinfo
  {year} {2013})}\BibitemShut {NoStop}%
\bibitem [{\citenamefont {Chen}\ \emph {et~al.}(2010)\citenamefont {Chen},
  \citenamefont {Gu},\ and\ \citenamefont {Wen}}]{Wen-SPT}%
  \BibitemOpen
  \bibfield  {author} {\bibinfo {author} {\bibfnamefont {X.}~\bibnamefont
  {Chen}}, \bibinfo {author} {\bibfnamefont {Z.-C.}\ \bibnamefont {Gu}}, \ and\
  \bibinfo {author} {\bibfnamefont {X.-G.}\ \bibnamefont {Wen}},\ }\href@noop
  {} {\bibfield  {journal} {\bibinfo  {journal} {Phys. Rev. B}\ }\textbf
  {\bibinfo {volume} {82}},\ \bibinfo {pages} {155138} (\bibinfo {year}
  {2010})}\BibitemShut {NoStop}%
\bibitem [{\citenamefont {Chen}\ \emph {et~al.}(2018)\citenamefont {Chen},
  \citenamefont {Li},\ and\ \citenamefont {Ng}}]{TKNg2018:prl}%
  \BibitemOpen
  \bibfield  {author} {\bibinfo {author} {\bibfnamefont {Z.}~\bibnamefont
  {Chen}}, \bibinfo {author} {\bibfnamefont {X.}~\bibnamefont {Li}}, \ and\
  \bibinfo {author} {\bibfnamefont {T.~K.}\ \bibnamefont {Ng}},\ }\href
  {\doibase 10.1103/PhysRevLett.120.046401} {\bibfield  {journal} {\bibinfo
  {journal} {Phys. Rev. Lett.}\ }\textbf {\bibinfo {volume} {120}},\ \bibinfo
  {pages} {046401} (\bibinfo {year} {2018})}\BibitemShut {NoStop}%
\bibitem [{\citenamefont {Li}\ \emph {et~al.}(2019)\citenamefont {Li},
  \citenamefont {Chen},\ and\ \citenamefont {Ng}}]{TKNg2019:prb}%
  \BibitemOpen
  \bibfield  {author} {\bibinfo {author} {\bibfnamefont {X.-H.}\ \bibnamefont
  {Li}}, \bibinfo {author} {\bibfnamefont {Z.}~\bibnamefont {Chen}}, \ and\
  \bibinfo {author} {\bibfnamefont {T.~K.}\ \bibnamefont {Ng}},\ }\href
  {\doibase 10.1103/PhysRevB.100.094519} {\bibfield  {journal} {\bibinfo
  {journal} {Phys. Rev. B}\ }\textbf {\bibinfo {volume} {100}},\ \bibinfo
  {pages} {094519} (\bibinfo {year} {2019})}\BibitemShut {NoStop}%
\bibitem [{\citenamefont {Wang}\ \emph {et~al.}(2017)\citenamefont {Wang},
  \citenamefont {Miao}, \citenamefont {Jin},\ and\ \citenamefont
  {Chen}}]{Miao2017:prb}%
  \BibitemOpen
  \bibfield  {author} {\bibinfo {author} {\bibfnamefont {Y.}~\bibnamefont
  {Wang}}, \bibinfo {author} {\bibfnamefont {J.-J.}\ \bibnamefont {Miao}},
  \bibinfo {author} {\bibfnamefont {H.-K.}\ \bibnamefont {Jin}}, \ and\
  \bibinfo {author} {\bibfnamefont {S.}~\bibnamefont {Chen}},\ }\href {\doibase
  10.1103/PhysRevB.96.205428} {\bibfield  {journal} {\bibinfo  {journal} {Phys.
  Rev. B}\ }\textbf {\bibinfo {volume} {96}},\ \bibinfo {pages} {205428}
  (\bibinfo {year} {2017})}\BibitemShut {NoStop}%
\bibitem [{\citenamefont {Miao}\ \emph {et~al.}(2017)\citenamefont {Miao},
  \citenamefont {Jin}, \citenamefont {Zhang},\ and\ \citenamefont
  {Zhou}}]{Miao2017:prl}%
  \BibitemOpen
  \bibfield  {author} {\bibinfo {author} {\bibfnamefont {J.-J.}\ \bibnamefont
  {Miao}}, \bibinfo {author} {\bibfnamefont {H.-K.}\ \bibnamefont {Jin}},
  \bibinfo {author} {\bibfnamefont {F.-C.}\ \bibnamefont {Zhang}}, \ and\
  \bibinfo {author} {\bibfnamefont {Y.}~\bibnamefont {Zhou}},\ }\href {\doibase
  10.1103/PhysRevLett.118.267701} {\bibfield  {journal} {\bibinfo  {journal}
  {Phys. Rev. Lett.}\ }\textbf {\bibinfo {volume} {118}},\ \bibinfo {pages}
  {267701} (\bibinfo {year} {2017})}\BibitemShut {NoStop}%
\bibitem [{\citenamefont {Miao}\ \emph {et~al.}(2019)\citenamefont {Miao},
  \citenamefont {Xu}, \citenamefont {Zhang},\ and\ \citenamefont
  {Zhang}}]{Miao2019:prb}%
  \BibitemOpen
  \bibfield  {author} {\bibinfo {author} {\bibfnamefont {J.-J.}\ \bibnamefont
  {Miao}}, \bibinfo {author} {\bibfnamefont {D.-H.}\ \bibnamefont {Xu}},
  \bibinfo {author} {\bibfnamefont {L.}~\bibnamefont {Zhang}}, \ and\ \bibinfo
  {author} {\bibfnamefont {F.-C.}\ \bibnamefont {Zhang}},\ }\href {\doibase
  10.1103/PhysRevB.99.245154} {\bibfield  {journal} {\bibinfo  {journal} {Phys.
  Rev. B}\ }\textbf {\bibinfo {volume} {99}},\ \bibinfo {pages} {245154}
  (\bibinfo {year} {2019})}\BibitemShut {NoStop}%
\bibitem [{\citenamefont {Ezawa}(2017)}]{Ezawa2017:prb}%
  \BibitemOpen
  \bibfield  {author} {\bibinfo {author} {\bibfnamefont {M.}~\bibnamefont
  {Ezawa}},\ }\href {\doibase 10.1103/PhysRevB.96.121105} {\bibfield  {journal}
  {\bibinfo  {journal} {Phys. Rev. B}\ }\textbf {\bibinfo {volume} {96}},\
  \bibinfo {pages} {121105} (\bibinfo {year} {2017})}\BibitemShut {NoStop}%
\bibitem [{\citenamefont {Ezawa}(2018)}]{Ezawa2018:prb}%
  \BibitemOpen
  \bibfield  {author} {\bibinfo {author} {\bibfnamefont {M.}~\bibnamefont
  {Ezawa}},\ }\href {\doibase 10.1103/PhysRevB.97.241113} {\bibfield  {journal}
  {\bibinfo  {journal} {Phys. Rev. B}\ }\textbf {\bibinfo {volume} {97}},\
  \bibinfo {pages} {241113} (\bibinfo {year} {2018})}\BibitemShut {NoStop}%
\bibitem [{\citenamefont {Kitaev}(2006)}]{Kitaev-model}%
  \BibitemOpen
  \bibfield  {author} {\bibinfo {author} {\bibfnamefont {A.}~\bibnamefont
  {Kitaev}},\ }\href@noop {} {\bibfield  {journal} {\bibinfo  {journal} {Ann.
  Phys.}\ }\textbf {\bibinfo {volume} {321}},\ \bibinfo {pages} {2} (\bibinfo
  {year} {2006})}\BibitemShut {NoStop}%
\bibitem [{\citenamefont {Shitade}\ \emph {et~al.}(2009)\citenamefont
  {Shitade}, \citenamefont {Katsura}, \citenamefont {Kuneš}, \citenamefont
  {Qi}, \citenamefont {Zhang},\ and\ \citenamefont
  {Nagaosa}}]{Shitade2009:prl}%
  \BibitemOpen
  \bibfield  {author} {\bibinfo {author} {\bibfnamefont {A.}~\bibnamefont
  {Shitade}}, \bibinfo {author} {\bibfnamefont {H.}~\bibnamefont {Katsura}},
  \bibinfo {author} {\bibfnamefont {J.}~\bibnamefont {Kuneš}}, \bibinfo
  {author} {\bibfnamefont {X.-L.}\ \bibnamefont {Qi}}, \bibinfo {author}
  {\bibfnamefont {S.-C.}\ \bibnamefont {Zhang}}, \ and\ \bibinfo {author}
  {\bibfnamefont {N.}~\bibnamefont {Nagaosa}},\ }\href {\doibase
  10.1103/PhysRevLett.102.256403} {\bibfield  {journal} {\bibinfo  {journal}
  {Phys. Rev. Lett.}\ }\textbf {\bibinfo {volume} {102}},\ \bibinfo {pages}
  {256403} (\bibinfo {year} {2009})}\BibitemShut {NoStop}%
\bibitem [{\citenamefont {R\"uegg}\ and\ \citenamefont
  {Fiete}(2012)}]{Ruegg2012:prl}%
  \BibitemOpen
  \bibfield  {author} {\bibinfo {author} {\bibfnamefont {A.}~\bibnamefont
  {R\"uegg}}\ and\ \bibinfo {author} {\bibfnamefont {G.~A.}\ \bibnamefont
  {Fiete}},\ }\href {\doibase 10.1103/PhysRevLett.108.046401} {\bibfield
  {journal} {\bibinfo  {journal} {Phys. Rev. Lett.}\ }\textbf {\bibinfo
  {volume} {108}},\ \bibinfo {pages} {046401} (\bibinfo {year}
  {2012})}\BibitemShut {NoStop}%
\bibitem [{\citenamefont {Rachel}(2018)}]{rachel2018:rpp}%
  \BibitemOpen
  \bibfield  {author} {\bibinfo {author} {\bibfnamefont {S.}~\bibnamefont
  {Rachel}},\ }\href@noop {} {\bibfield  {journal} {\bibinfo  {journal}
  {Reports on Progress in Physics}\ }\textbf {\bibinfo {volume} {81}},\
  \bibinfo {pages} {116501} (\bibinfo {year} {2018})}\BibitemShut {NoStop}%
\bibitem [{\citenamefont {Kimchi}(2015)}]{kimchi2015:phd}%
  \BibitemOpen
  \bibfield  {author} {\bibinfo {author} {\bibfnamefont {I.}~\bibnamefont
  {Kimchi}},\ }\emph {\bibinfo {title} {Spin-Orbit Coupled Quantum Magnetism in
  the 3D-Honeycomb Iridates}},\ \href@noop {} {Ph.D. thesis},\ \bibinfo
  {school} {University of California, Berkeley} (\bibinfo {year}
  {2015})\BibitemShut {NoStop}%
\bibitem [{\citenamefont {Fu}\ and\ \citenamefont {Kane}(2008)}]{FK2008:prl}%
  \BibitemOpen
  \bibfield  {author} {\bibinfo {author} {\bibfnamefont {L.}~\bibnamefont
  {Fu}}\ and\ \bibinfo {author} {\bibfnamefont {C.~L.}\ \bibnamefont {Kane}},\
  }\href {\doibase 10.1103/PhysRevLett.100.096407} {\bibfield  {journal}
  {\bibinfo  {journal} {Phys. Rev. Lett.}\ }\textbf {\bibinfo {volume} {100}},\
  \bibinfo {pages} {096407} (\bibinfo {year} {2008})}\BibitemShut {NoStop}%
\bibitem [{\citenamefont {Xu}\ \emph {et~al.}(2014)\citenamefont {Xu},
  \citenamefont {Liu}, \citenamefont {Wang}, \citenamefont {Ge}, \citenamefont
  {Liu}, \citenamefont {Yang}, \citenamefont {Chen}, \citenamefont {Liu},
  \citenamefont {Xu}, \citenamefont {Gao}, \citenamefont {Qian}, \citenamefont
  {Zhang},\ and\ \citenamefont {Jia}}]{Xu2014:prl}%
  \BibitemOpen
  \bibfield  {author} {\bibinfo {author} {\bibfnamefont {J.-P.}\ \bibnamefont
  {Xu}}, \bibinfo {author} {\bibfnamefont {C.}~\bibnamefont {Liu}}, \bibinfo
  {author} {\bibfnamefont {M.-X.}\ \bibnamefont {Wang}}, \bibinfo {author}
  {\bibfnamefont {J.}~\bibnamefont {Ge}}, \bibinfo {author} {\bibfnamefont
  {Z.-L.}\ \bibnamefont {Liu}}, \bibinfo {author} {\bibfnamefont
  {X.}~\bibnamefont {Yang}}, \bibinfo {author} {\bibfnamefont {Y.}~\bibnamefont
  {Chen}}, \bibinfo {author} {\bibfnamefont {Y.}~\bibnamefont {Liu}}, \bibinfo
  {author} {\bibfnamefont {Z.-A.}\ \bibnamefont {Xu}}, \bibinfo {author}
  {\bibfnamefont {C.-L.}\ \bibnamefont {Gao}}, \bibinfo {author} {\bibfnamefont
  {D.}~\bibnamefont {Qian}}, \bibinfo {author} {\bibfnamefont {F.-C.}\
  \bibnamefont {Zhang}}, \ and\ \bibinfo {author} {\bibfnamefont {J.-F.}\
  \bibnamefont {Jia}},\ }\href {\doibase 10.1103/PhysRevLett.112.217001}
  {\bibfield  {journal} {\bibinfo  {journal} {Phys. Rev. Lett.}\ }\textbf
  {\bibinfo {volume} {112}},\ \bibinfo {pages} {217001} (\bibinfo {year}
  {2014})}\BibitemShut {NoStop}%
\bibitem [{\citenamefont {Sau}\ \emph {et~al.}(2010)\citenamefont {Sau},
  \citenamefont {Lutchyn}, \citenamefont {Tewari},\ and\ \citenamefont
  {Das~Sarma}}]{Sau2010:prb}%
  \BibitemOpen
  \bibfield  {author} {\bibinfo {author} {\bibfnamefont {J.~D.}\ \bibnamefont
  {Sau}}, \bibinfo {author} {\bibfnamefont {R.~M.}\ \bibnamefont {Lutchyn}},
  \bibinfo {author} {\bibfnamefont {S.}~\bibnamefont {Tewari}}, \ and\ \bibinfo
  {author} {\bibfnamefont {S.}~\bibnamefont {Das~Sarma}},\ }\href {\doibase
  10.1103/PhysRevB.82.094522} {\bibfield  {journal} {\bibinfo  {journal} {Phys.
  Rev. B}\ }\textbf {\bibinfo {volume} {82}},\ \bibinfo {pages} {094522}
  (\bibinfo {year} {2010})}\BibitemShut {NoStop}%
\bibitem [{\citenamefont {Stanescu}\ \emph {et~al.}(2010)\citenamefont
  {Stanescu}, \citenamefont {Sau}, \citenamefont {Lutchyn},\ and\ \citenamefont
  {Das~Sarma}}]{Stanescu2010:prb}%
  \BibitemOpen
  \bibfield  {author} {\bibinfo {author} {\bibfnamefont {T.~D.}\ \bibnamefont
  {Stanescu}}, \bibinfo {author} {\bibfnamefont {J.~D.}\ \bibnamefont {Sau}},
  \bibinfo {author} {\bibfnamefont {R.~M.}\ \bibnamefont {Lutchyn}}, \ and\
  \bibinfo {author} {\bibfnamefont {S.}~\bibnamefont {Das~Sarma}},\ }\href
  {\doibase 10.1103/PhysRevB.81.241310} {\bibfield  {journal} {\bibinfo
  {journal} {Phys. Rev. B}\ }\textbf {\bibinfo {volume} {81}},\ \bibinfo
  {pages} {241310} (\bibinfo {year} {2010})}\BibitemShut {NoStop}%
\bibitem [{\citenamefont {Lababidi}\ and\ \citenamefont
  {Zhao}(2011)}]{Lababidi2011:prb}%
  \BibitemOpen
  \bibfield  {author} {\bibinfo {author} {\bibfnamefont {M.}~\bibnamefont
  {Lababidi}}\ and\ \bibinfo {author} {\bibfnamefont {E.}~\bibnamefont
  {Zhao}},\ }\href {\doibase 10.1103/PhysRevB.83.184511} {\bibfield  {journal}
  {\bibinfo  {journal} {Phys. Rev. B}\ }\textbf {\bibinfo {volume} {83}},\
  \bibinfo {pages} {184511} (\bibinfo {year} {2011})}\BibitemShut {NoStop}%
\bibitem [{\citenamefont {Chen}\ and\ \citenamefont
  {Franz}(2016)}]{Chen2016:prb}%
  \BibitemOpen
  \bibfield  {author} {\bibinfo {author} {\bibfnamefont {A.}~\bibnamefont
  {Chen}}\ and\ \bibinfo {author} {\bibfnamefont {M.}~\bibnamefont {Franz}},\
  }\href {\doibase 10.1103/PhysRevB.93.201105} {\bibfield  {journal} {\bibinfo
  {journal} {Phys. Rev. B}\ }\textbf {\bibinfo {volume} {93}},\ \bibinfo
  {pages} {201105} (\bibinfo {year} {2016})}\BibitemShut {NoStop}%
\bibitem [{\citenamefont {Schnyder}\ \emph
  {et~al.}(2008{\natexlab{b}})\citenamefont {Schnyder}, \citenamefont {Ryu},
  \citenamefont {Furusaki},\ and\ \citenamefont {Ludwig}}]{Ryu2008:prb}%
  \BibitemOpen
  \bibfield  {author} {\bibinfo {author} {\bibfnamefont {A.~P.}\ \bibnamefont
  {Schnyder}}, \bibinfo {author} {\bibfnamefont {S.}~\bibnamefont {Ryu}},
  \bibinfo {author} {\bibfnamefont {A.}~\bibnamefont {Furusaki}}, \ and\
  \bibinfo {author} {\bibfnamefont {A.~W.~W.}\ \bibnamefont {Ludwig}},\ }\href
  {\doibase 10.1103/PhysRevB.78.195125} {\bibfield  {journal} {\bibinfo
  {journal} {Phys. Rev. B}\ }\textbf {\bibinfo {volume} {78}},\ \bibinfo
  {pages} {195125} (\bibinfo {year} {2008}{\natexlab{b}})}\BibitemShut
  {NoStop}%
\bibitem [{\citenamefont {Qi}\ \emph {et~al.}(2009)\citenamefont {Qi},
  \citenamefont {Hughes}, \citenamefont {Raghu},\ and\ \citenamefont
  {Zhang}}]{Qi2009:prl}%
  \BibitemOpen
  \bibfield  {author} {\bibinfo {author} {\bibfnamefont {X.-L.}\ \bibnamefont
  {Qi}}, \bibinfo {author} {\bibfnamefont {T.~L.}\ \bibnamefont {Hughes}},
  \bibinfo {author} {\bibfnamefont {S.}~\bibnamefont {Raghu}}, \ and\ \bibinfo
  {author} {\bibfnamefont {S.-C.}\ \bibnamefont {Zhang}},\ }\href {\doibase
  10.1103/PhysRevLett.102.187001} {\bibfield  {journal} {\bibinfo  {journal}
  {Phys. Rev. Lett.}\ }\textbf {\bibinfo {volume} {102}},\ \bibinfo {pages}
  {187001} (\bibinfo {year} {2009})}\BibitemShut {NoStop}%
\bibitem [{\citenamefont {Sheng}\ \emph {et~al.}(2005)\citenamefont {Sheng},
  \citenamefont {Sheng}, \citenamefont {Ting},\ and\ \citenamefont
  {Haldane}}]{Sheng2005:prl}%
  \BibitemOpen
  \bibfield  {author} {\bibinfo {author} {\bibfnamefont {L.}~\bibnamefont
  {Sheng}}, \bibinfo {author} {\bibfnamefont {D.~N.}\ \bibnamefont {Sheng}},
  \bibinfo {author} {\bibfnamefont {C.~S.}\ \bibnamefont {Ting}}, \ and\
  \bibinfo {author} {\bibfnamefont {F.~D.~M.}\ \bibnamefont {Haldane}},\ }\href
  {\doibase 10.1103/PhysRevLett.95.136602} {\bibfield  {journal} {\bibinfo
  {journal} {Phys. Rev. Lett.}\ }\textbf {\bibinfo {volume} {95}},\ \bibinfo
  {pages} {136602} (\bibinfo {year} {2005})}\BibitemShut {NoStop}%
\bibitem [{\citenamefont {Sheng}\ \emph {et~al.}(2006)\citenamefont {Sheng},
  \citenamefont {Weng}, \citenamefont {Sheng},\ and\ \citenamefont
  {Haldane}}]{Sheng2006:prl}%
  \BibitemOpen
  \bibfield  {author} {\bibinfo {author} {\bibfnamefont {D.~N.}\ \bibnamefont
  {Sheng}}, \bibinfo {author} {\bibfnamefont {Z.~Y.}\ \bibnamefont {Weng}},
  \bibinfo {author} {\bibfnamefont {L.}~\bibnamefont {Sheng}}, \ and\ \bibinfo
  {author} {\bibfnamefont {F.~D.~M.}\ \bibnamefont {Haldane}},\ }\href
  {\doibase 10.1103/PhysRevLett.97.036808} {\bibfield  {journal} {\bibinfo
  {journal} {Phys. Rev. Lett.}\ }\textbf {\bibinfo {volume} {97}},\ \bibinfo
  {pages} {036808} (\bibinfo {year} {2006})}\BibitemShut {NoStop}%
\bibitem [{\citenamefont {Prodan}(2009)}]{Prodan2009:prb}%
  \BibitemOpen
  \bibfield  {author} {\bibinfo {author} {\bibfnamefont {E.}~\bibnamefont
  {Prodan}},\ }\href {\doibase 10.1103/PhysRevB.80.125327} {\bibfield
  {journal} {\bibinfo  {journal} {Phys. Rev. B}\ }\textbf {\bibinfo {volume}
  {80}},\ \bibinfo {pages} {125327} (\bibinfo {year} {2009})}\BibitemShut
  {NoStop}%
\bibitem [{\citenamefont {Rachel}(2016)}]{rachel2016:jpc}%
  \BibitemOpen
  \bibfield  {author} {\bibinfo {author} {\bibfnamefont {S.}~\bibnamefont
  {Rachel}},\ }\href@noop {} {\bibfield  {journal} {\bibinfo  {journal}
  {Journal of Physics: Condensed Matter}\ }\textbf {\bibinfo {volume} {28}},\
  \bibinfo {pages} {405502} (\bibinfo {year} {2016})}\BibitemShut {NoStop}%
\bibitem [{\citenamefont {Ezawa}\ \emph {et~al.}(2013)\citenamefont {Ezawa},
  \citenamefont {Tanaka},\ and\ \citenamefont {Nagaosa}}]{ezawa2013:sr}%
  \BibitemOpen
  \bibfield  {author} {\bibinfo {author} {\bibfnamefont {M.}~\bibnamefont
  {Ezawa}}, \bibinfo {author} {\bibfnamefont {Y.}~\bibnamefont {Tanaka}}, \
  and\ \bibinfo {author} {\bibfnamefont {N.}~\bibnamefont {Nagaosa}},\
  }\href@noop {} {\bibfield  {journal} {\bibinfo  {journal} {Scientific
  reports}\ }\textbf {\bibinfo {volume} {3}},\ \bibinfo {pages} {2790}
  (\bibinfo {year} {2013})}\BibitemShut {NoStop}%
\bibitem [{\citenamefont {Yang}\ \emph {et~al.}(2011)\citenamefont {Yang},
  \citenamefont {Xu}, \citenamefont {Sheng}, \citenamefont {Wang},
  \citenamefont {Xing},\ and\ \citenamefont {Sheng}}]{Sheng2011:prl}%
  \BibitemOpen
  \bibfield  {author} {\bibinfo {author} {\bibfnamefont {Y.}~\bibnamefont
  {Yang}}, \bibinfo {author} {\bibfnamefont {Z.}~\bibnamefont {Xu}}, \bibinfo
  {author} {\bibfnamefont {L.}~\bibnamefont {Sheng}}, \bibinfo {author}
  {\bibfnamefont {B.}~\bibnamefont {Wang}}, \bibinfo {author} {\bibfnamefont
  {D.~Y.}\ \bibnamefont {Xing}}, \ and\ \bibinfo {author} {\bibfnamefont
  {D.~N.}\ \bibnamefont {Sheng}},\ }\href {\doibase
  10.1103/PhysRevLett.107.066602} {\bibfield  {journal} {\bibinfo  {journal}
  {Phys. Rev. Lett.}\ }\textbf {\bibinfo {volume} {107}},\ \bibinfo {pages}
  {066602} (\bibinfo {year} {2011})}\BibitemShut {NoStop}%
\bibitem [{\citenamefont {Li}\ \emph {et~al.}(2013)\citenamefont {Li},
  \citenamefont {Hui-Chao}, \citenamefont {Yun-You}, \citenamefont
  {Dong-Ning},\ and\ \citenamefont {Ding-Yu}}]{Sheng2013:cpb}%
  \BibitemOpen
  \bibfield  {author} {\bibinfo {author} {\bibfnamefont {S.}~\bibnamefont
  {Li}}, \bibinfo {author} {\bibfnamefont {L.}~\bibnamefont {Hui-Chao}},
  \bibinfo {author} {\bibfnamefont {Y.}~\bibnamefont {Yun-You}}, \bibinfo
  {author} {\bibfnamefont {S.}~\bibnamefont {Dong-Ning}}, \ and\ \bibinfo
  {author} {\bibfnamefont {X.}~\bibnamefont {Ding-Yu}},\ }\href {\doibase
  doi.org/10.1088/1674-1056/22/6/067201} {\bibfield  {journal} {\bibinfo
  {journal} {Chinese Physics B}\ }\textbf {\bibinfo {volume} {22}},\ \bibinfo
  {pages} {067201} (\bibinfo {year} {2013})}\BibitemShut {NoStop}%
\end{thebibliography}%

\end{document}